\documentclass[aps,prl,twocolumn,showpacs,superscriptaddress,longbibliography]{revtex4-2}
\usepackage{amsfonts}
\usepackage{amsmath}
\usepackage{graphicx}
\usepackage{bm}
\usepackage{soul}
\usepackage{amssymb}
\usepackage{dcolumn}
\usepackage{color}
\usepackage{multirow}
\usepackage{booktabs}
\usepackage{latexsym}
\usepackage[colorlinks,
linkcolor=blue,
anchorcolor=blue,
citecolor=blue,
urlcolor=blue,
]{hyperref}

\begin{document}
    \title{Intercalation-Induced Near Room-Temperature Ferromagnetism in CrI$_3$ via Synergistic Exchange Pathways}

    \author{Qing-Han Yang}
    \affiliation{Kavli Institute for Theoretical Sciences, University of Chinese Academy of Sciences, Beijing 100049, China}

    \author{Jia-Wen Li}
    \affiliation{Kavli Institute for Theoretical Sciences, University of Chinese Academy of Sciences, Beijing 100049, China}

    \author{Xin-Wei Yi}
    \affiliation{School of Physical Sciences, University of Chinese Academy of Sciences, Beijing 100049, China}

    \author{Xiang Li}
    \affiliation{Kavli Institute for Theoretical Sciences, University of Chinese Academy of Sciences, Beijing 100049, China}

    \author{Jing-Yang You}
    \affiliation{Peng Huanwu Collaborative Center for Research and Education, Beihang University, Beijing 100191, China}

    \author{Gang Su}
    \email{gsu@ucas.ac.cn}
    \affiliation{Kavli Institute for Theoretical Sciences, University of Chinese Academy of Sciences, Beijing 100049, China}
    \affiliation{School of Physical Sciences, University of Chinese Academy of Sciences, Beijing 100049, China}
    \affiliation{Institute of Theoretical Physics, Chinese Academy of Sciences, Beijing 100190, China}
    \affiliation{Physical Science Laboratory, Huairou National Comprehensive Science Center, Beijing 101400, China}

    \author{Bo Gu}
    \email{gubo@ucas.ac.cn}
    \affiliation{Kavli Institute for Theoretical Sciences, University of Chinese Academy of Sciences, Beijing 100049, China}
    \affiliation{Physical Science Laboratory, Huairou National Comprehensive Science Center, Beijing 101400, China}

    \begin{abstract}
        The development of room-temperature magnetic semiconductors is critical for advancing spintronic technologies, 
        yet van der Waals magnets like CrI$_3$ exhibit intrinsically low Curie temperatures ($T{\rm{_C}}$ $\sim$ 45 K). 
        This study employs first-principles calculations to demonstrate that atom intercalation, particularly lithium 
        (Li), dramatically enhances magnetic exchange couplings in CrI$_3$, achieving near room-temperature ferromagnetism 
        with a predicted $T{\rm{_C}}$ of 286 K-aligning with experimental reports of 420 K. The underlying mechanism 
        involves synergistic superexchange and double-exchange interactions: intercalation reduces the $|E_p-E_d|$ 
        energy difference between iodine p-orbitals and chromium d-orbitals, strengthening superexchange pathways, 
        while charge transfer induces valence mixing (e.g., Cr$^{3+}$ to Cr$^{2+}$, as confirmed by experimental 
        X-ray photoelectron spectrometry data), promoting double-exchange. Theoretical predictions extend to other 
        intercalants including Cu and Na, with Cu$_{0.25}$CrI$_3$ and Na$_{0.25}$CrI$_3$ 
        exhibiting $T{\rm{_C}}$ of 267 K and 247 K, respectively, establishing a versatile 
        strategy for designing high-$T{\rm{_C}}$ magnetic semiconductors. This work bridges theoretical insights 
        with experimental validation, offering a transferable framework for intercalation-driven material design 
        and accelerating practical spintronic device realization. 
    \end{abstract}
    \maketitle

    %%%%%%% Main text %%%%%%%%%%%%%%%%%%%%%
    \par {\it \textcolor{blue}{Introduction.}}---Room-temperature (RT) magnetic semiconductors are crucial for advancing 
    spintronic, allowing for the control of both electron spin and charge. This capability promises next-generation 
    devices with enhanced functionality and energy efficiency. Recent breakthroughs in intrinsic ferromagnetism within 
    two-dimensional (2D) van der Waals (vdW) materials—exemplified by CrI$_3$ \cite{huangCrI32017}, CrSBr 
    \cite{lee2021}, and Cr$_2$Ge$_2$Te$_6$ \cite{gong2017}—have established new platforms 
    for fundamental studies. However, their Curie temperatures (e.g., $T{\rm{_C}}$ $\approx$ 45 K for monolayer CrI$_3$ 
    \cite{huangCrI32017}) remain substantially below the operational threshold for practical applications. 
    This critical temperature limitation necessitates innovative strategies to engineer vdW magnets with RT 
    magnetic order—a prerequisite for advancing both fundamental research and scalable spintronic implementations. 

    \par Extensive research confirms that 2D material properties are tunable via external stimuli—including 
    electric fields \cite{dengGate2018,jiangElectric2018}, doping \cite{el-habib2024,jiangdoping2018}, 
    strain \cite{nistrain2021,oneill2023,zeng,youP2023}, surface functionalization \cite{datta2017,gonzalez2016}, and 
    intercalation \cite{rajapakse2021,wang2024,zhang2025,liu2024,kappera2014,yangReview2022,yangReview2023,yangReview2024a,
    zhu2019,kanetani2012,xiongLi2015,yaoCu2014,zhang2022,feuer2025,huanFe2023,iturriaga2023,liu2024,mishra2024,mi2022,
    weber2019,wuFe2023,zhaoCu2024,bensch2009,ji2021,xuLi2024,huempfner2022,huempfner2023,kim2024,liK2024,liuCu2017,
    nongCu2025,scholzel2024,zhangK2020,tonti2000,toyama2022,wangNaAl2015,naik2021,meng2020}. 
    Among these, intercalation (insertion of atoms, ions, or molecules into host layers) has emerged as a 
    pivotal technique for engineering vdW materials, generating synergistic effects that modify exchange coupling 
    interactions \cite{wang2024,zhang2025}, charge transfer dynamics \cite{liu2024,wang2024}, in-plane bonding configurations 
    \cite{kappera2014,yangReview2022,yangReview2023,zhu2019}, electronic band structures \cite{kanetani2012,xiongLi2015,yaoCu2014,zhang2022}, 
    and spin-orbit effects \cite{huangLi2021,zhaosoc2020}. 
    %These alterations enhance photonic, electronic, optoelectronic, thermoelectric, and magnetic properties. 
    Crucially, intercalation can induce ferromagnetic order or substantially 
    elevate critical temperatures in non-magnetic or low-$T{\rm{_C}}$ layered systems 
    \cite{feuer2025,huanFe2023,iturriaga2023,liu2024,mishra2024,mi2022,wang2024,weber2019,wuFe2023,zhang2025,zhaoCu2024}. 
    %as demonstrated by RT ferromagnetism in ammonium-intercalated VOCl \cite{liu2024}, Cu-intercalated WO$_3$ 
    %nanosheets \cite{zhaoCu2024}, and Fe-self-intercalated FeSe$_2$ \cite{huanFe2023}. 
    Lithium (Li) represents a highly prominent effective intercalant for synthesizing 2D nanosheets and modulating 
    diverse properties \cite{bensch2009,feuer2025,ji2021,kappera2014,wang2024,xiongLi2015,xuLi2024,zhu2019}. Recent 
    experiments report RT ferromagnetic order in Li-intercalated CrI$_3$ (Li$_n$CrI$_3$), where 
    $T{\rm{_C}}$ increases from 45 K to 420 K while preserving semiconducting behavior \cite{wang2024}, motivating 
    our investigation into the underlying mechanism of Li-intercalation-induced ferromagnetism. 
    
    %including electrochemical energy storage, RT magnetic materials
    %The antiferromagnetic (AFM)-ferrimagnetic (FIM)-AFM transition can be obtained in 
    %NiPS$_3$ by intercalating organic cations into the vdW gaps to provide electron doping \cite{}. 

    \begin{figure}[tphb]
    	\centering
        \includegraphics[scale=0.6]{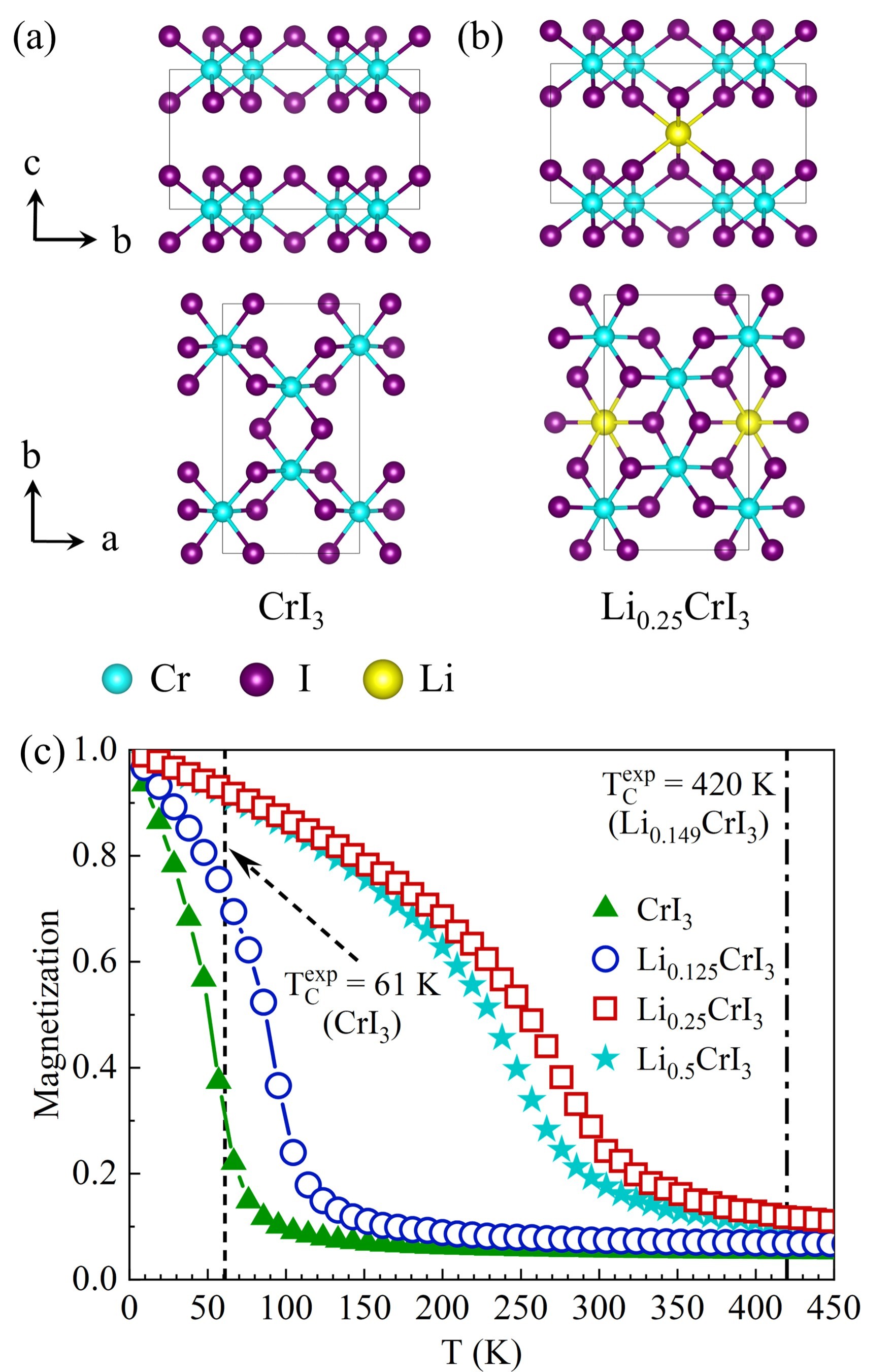}
        \caption{
    	The crystal structures of (a) HT phase CrI$_3$ and (b) Li$_{0.25}$CrI$_3$. (c) Comparsion of experimental 
        and calculated $T{\rm{_C}}$ values for Li-intercalated CrI$_3$, obtained from Monte Carlo (MC) simulations. 
    	}\label{Fig1}
    \end{figure}

    \par In this work, we conduct first-principles calculations to elucidate the mechanism of Li-intercalation-induced 
    RT ferromagnetic (FM) order in 2D CrI$_3$, examining diverse intercalation sites and Li nominal 
    stoichiometries (n = 0.125, 0.25, 0.5). Our results demonstrate that Li intercalation substantially enhances both 
    intralayer and interlayer exchange couplings, leading to a marked increase in $T{\rm{_C}}$ from 57 K to 286 K—a 
    trend qualitatively aligned with experimental reports. The $T{\rm{_C}}$ enhancement primarily arises from synergistic 
    superexchange and double-exchange interactions, attributed to reduced energy-level differences between Cr d and I p 
    orbitals. Extending our analysis to other intercalants at a fixed stoichiometry of 0.25, we 
    identify Cu$_{0.25}$CrI$_3$ and Na$_{0.25}$CrI$_3$ as systems exhibiting near RT FM order, 
    underscoring intercalation's predictive efficacy for alkali metals in elevating $T{\rm{_C}}$. This study establishes a 
    robust theoretical foundation for achieving near RT magnetism via intercalation and pinpoints experimentally accessible 
    materials for next-generation spintronic devices. 

    \begin{table}[!htbp]
        \centering
        \caption{Calculated magnetic exchange integrals ($J_1$ (meV), $J_2$ (meV), and $J_1'$ (meV)), $T{\rm{_C}}$ (K), 
        and gap (eV) for CrI$_3$ and Li$_n$CrI$_3$ ($n$ = 0.125, 0.25, 0.5). }
        \vspace{3pt}
        \label{table1}
        \setlength{\tabcolsep}{0.24cm}
        \renewcommand{\arraystretch}{1.2}
        \begin{tabular}{cccccc}
            \toprule
            & $J_1$ & $J_2$ & $J_1'$ & $T{\rm{_C}}$ & Gap \\
            \midrule
            CrI$_3$ & -6.344 & -0.581 & -2.685 & 57 & 1.16 \\
            CrI$_3$ (HT) & -6.090 & -0.523 & 0.047 & 47 ($T{\rm{_N}}$) & 1.24 \\
            Li$_{0.125}$CrI$_3$ & -6.946 & -1.052 & -3.070 & 95 & 0.03 \\
            Li$_{0.25}$CrI$_3$ & -17.335 & -5.388 & -2.836 & 286 & 0.05 \\
            Li$_{0.5}$CrI$_3$ & -13.387 & -4.061 & -9.623 & 257 & 0 \\
            \bottomrule
        \end{tabular}
    \end{table}

    \par {\it \textcolor{blue}{$J_1$-$J_2$-$J_1'$ Heisenberg model.}}---Bulk CrI$_3$ experimentally exhibits two distinct 
    structural phases: rhombohedral ($R\bar{3}$) at low temperature (LT) and monoclinic ($C2/m$) above 210-220 K 
    (HT phase) \cite{mcguire2015}. The structure of HT phase CrI$_3$ is illustrated in Fig. \ref{Fig1}(a). 
    Raman spectroscopy confirms the persistence of monoclinic structure in thin samples down to cryogenic 
    temperature \cite{ubrig2019}, justifying our selection of the HT phase for intercalation studies. 

    \begin{figure*}[tphb]
    	\centering
        \includegraphics[scale=0.52]{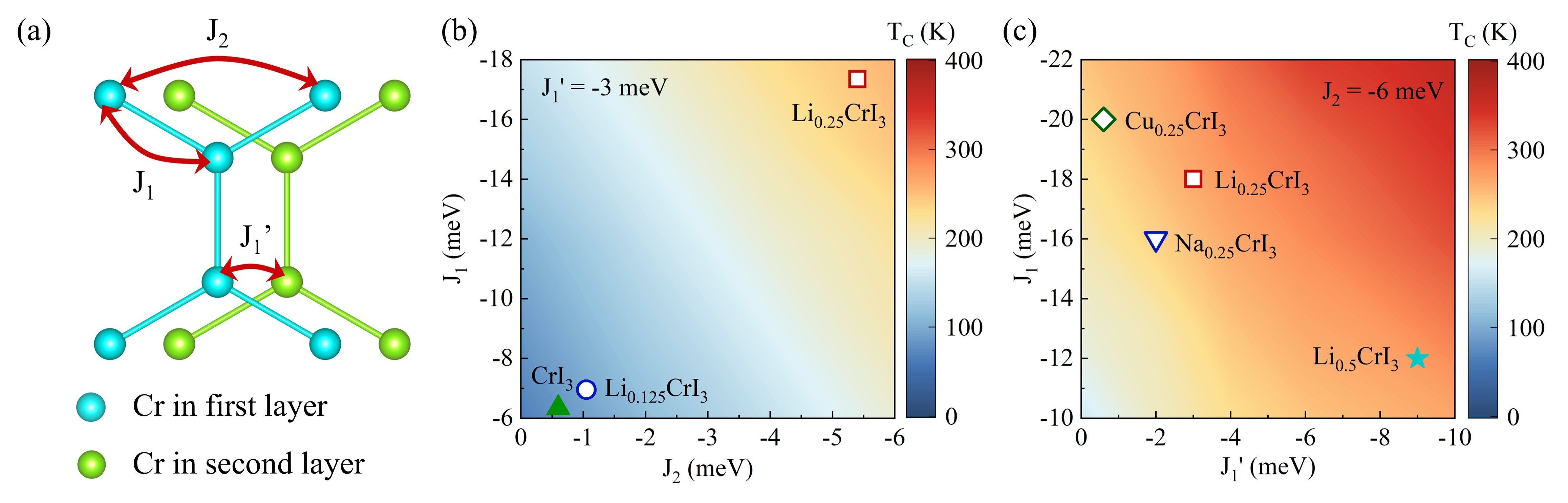}
        \caption{
            (a) Schematic diagram illustrates $J_1$, $J_2$ and $J_1'$. The phase diagrams are presented for (b) intralayer 
            nearest-neighbor ($J_1$) and next nearest-neighbor ($J_2$) exchange integrals, with $J_1'$ fixed at -3 meV; and 
            (c) intralayer ($J_1$) and interlayer ($J_1'$) nearest-neighbor exchange integrals, with $J_2$ fixed at -6 meV. 
    	}\label{Fig2}
    \end{figure*}

    \par To investigate magnetic properties of CrI$_3$, we considered the following Heisenberg-type Hamiltonian: 

    \begin{equation}
        \begin{split}
            H =& \sum_{<i,j>} J_1 \textbf{S}_{i} \cdot \textbf{S}_{j} + \sum_{<<i,j>>} J_2 \textbf{S}_{i} \cdot \textbf{S}_{j} + 
            \sum_{<i,i'>} J_1' \textbf{S}_{i} \cdot \textbf{S}_{i'} \\
            &+ A \sum_{i} (S^z_i)^2 + E_0, 
        \end{split}
        \label{equ1}
    \end{equation}

    \noindent where $J_1$ and $J_2$ denote the nearest-neighbor (NN) and next-nearest-neighbor (NNN) intralayer exchange couplings, 
    respectively, while $J_1'$ represents the NN interlayer exchange coupling. For notational simplicity, $J_1$ is used in place of 
    $J_1 \textbf{S}^2$ throughout this work, with the same convention applying to $J_2$ and $J_1'$. These couplings are illustrated 
    in Fig. \ref{Fig2}(a). $A$ is the magnetic anisotropy energy (MAE, defined as $E_\parallel - E_\perp$). The exchange coupling 
    parameters were determined by 
    evaluating four distinct spin configurations (detailed in Supplementary Materials (SMs) \cite{SMs}). Our calculations show a 
    $T{\rm{_C}}$ of 57 K for the LT phase, consistent with experimental values \cite{mcguire2015}. For the HT phase, we observe 
    antiferromagnetic (AFM) interlayer coupling with a calculated Néel temperature ($T{\rm{_N}}$) of 47 K, matching experimental 
    reports of 51 K critical temperature with AFM order \cite{wangAFM2018}. 

    \par To elucidate the origin of the enhanced $T{\rm{_C}}$ observed experimentally, we systematically investigated five 
    high-symmetry intercalation sites (detailed in SMs \cite{SMs}). The experimental value is 0.15. Since direct calculation of 
    the supercell at this value would incur a large computational burden, we substituted values of 0.125, 0.25 and 0.5. 
    %The relative energies of these five sites, presented in Fig. S1, identify Site 5 as the most energetically stable configuration, 
    %while Site 4 corresponds to a metastable state. 
    As shown in Table. S1, Site 4 (Li$_{0.25}$CrI$_3$) exhibits the highest calculated $T{\rm{_C}}$ of 286 K, closely matching 
    experimental reports, and was chosen for further study. 
    %Structural characterization reveals that Li$_{0.25}$CrI$_3$ adopts lattice parameter $a = 6.98$ {\AA}, $b = 12.30$ {\AA}, 
    %and $c = 6.73$ {\AA}, representing a contraction along the $c$-axis ($\Delta c$ = -0.39 {\AA}, -5.5\% vs. HT phase 
    %$c = 7.12$ {\AA}) with minimal variation along $a$ ($\Delta a$ = +0.01 {\AA}) and $b$ ($\Delta b$ = +0.23 {\AA}) relative 
    %to the HT phase ($a = 6.97$ {\AA}, $b = 12.07$ {\AA}). 
    This intercalation accompanies a space group transition from $C2/m$ to $P2/m$, as visualized in the crystal structure 
    comparison. Concurrently, Li intercalation induces a magnetic reconfiguration: the interlayer coupling switches from AFM 
    to FM, while the MAE increases significantly (Table. S2) with the easy axis reoriented from out-of-plane to in-plane. 
    %Exchange coupling parameters for Li$_{0.25}$CrI$_3$ are compiled in Table \ref{table1}. 
    We computed $T{\rm{_C}}$ for Li$_n$CrI$_3$ (n = 0.125, 0.25, and 0.5), guided by experimentally achievable Li 
    stoichiometries \cite{wang2024}, revealing an increase from $\sim$ 57 K to 286 K with rising Li content (Table. \ref{table1}). 
    This trend aligns qualitatively with experimental measurements \cite{wang2024}. The discrepancy in $T{\rm{_C}}$ 
    may arise from stoichiometric fluctuations or interlayer stacking disorders, as observed in Ref \cite{wang2024}. 
    %where $T{\rm{_C}}$ enhancement correlates with increased Li stoichiometry 

    \par To quantitatively elucidate the influence of exchange couplings ($J_1$, $J_2$, and $J_1'$) on $T{\rm{_C}}$, we systematically 
    mapped the parameter space by holding $J_1'$ constant at -3 meV while varying $J_1$ and $J_2$, and separately fixing $J_2$ at 
    -6 meV while modulating $J_1$ and $J_1'$. The resulting phase diagrams, presented in Fig. \ref{Fig2}(b) and (c), reveal distinct 
    magnetic regimes with explicit annotation of Li-intercalated CrI$_3$ data points. Comparative analysis demonstrates that 25\% Li 
    intercalation (Li$_{0.25}$CrI$_3$) enhances both $J_1$ and $J_2$ relative to pristine CrI$_3$, indicating 
    generalized strengthening of intralayer exchange interactions. However, reducing Li stoichiometry to 0.125 lowers $J_1$ and $J_2$, 
    while $J_1'$ remains invariant, resulting in a lower $T{\rm{_C}}$. Conversely, at 0.5 stoichiometry, $J_1$ decreases while $J_1'$ 
    increases significantly and $J_2$ remains stable. This compensation leads to an unchanged $T{\rm{_C}}$. 
    %This stoichiometric dependence establishes that $T{\rm{_C}}$ optimization primarily correlates 
    %with intralayer coupling enhancement rather than interlayer interactions. 

    \begin{figure*}[!tphb]
    	\centering
        \includegraphics[scale=0.5]{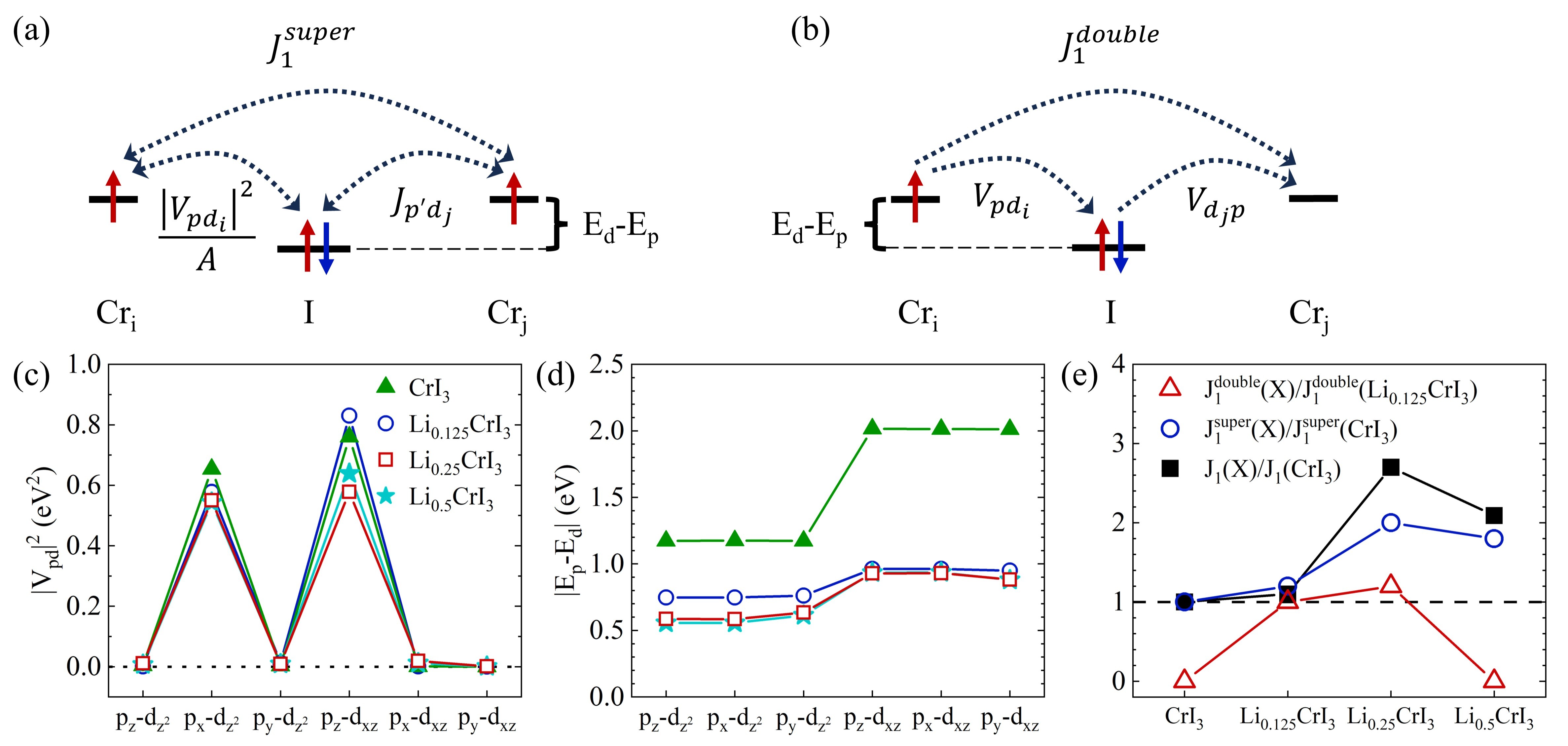}
        \caption{
    	The diagram of (a) superexchange interaction and (b) double-exchange interaction. (c) Hopping parameters $|V_{pd}|^2$ and (d) 
        energy difference $|E_p - E_d|$ between $p$ orbitals of I atoms and $d_{z^2}$, $d_{xz}$ orbitals of Cr atoms. (e) Ratios of 
        the nearest-neighbor exchange coupling for DFT-calculated ($J_1$ (X)/$J_1$ (CrI$_3$)) in black line, superexchange 
        model-calculated ($J^{super}_1$ (X)/$J^{super}_1$ (CrI$_3$)) in blue line, and double-exchange model-calculated 
        ($J^{double}_1$ (X)/$J^{double}_1$ (Li$_{0.125}$CrI$_3$)) in red line. Here, X represents CrI$_3$, Li$_{0.125}$CrI$_3$, 
        Li$_{0.25}$CrI$_3$, and Li$_{0.5}$CrI$_3$. The data is listed in Table. \ref{table2} and Table. \ref{table3}. 
    	}\label{Fig3}
    \end{figure*}

    %\begin{table*}
    %    \centering
    %    \caption{1}
    %    \vspace{5pt}
    %    \setlength{\tabcolsep}{0.4cm}
    %    \renewcommand{\arraystretch}{1.5}
    %    \begin{tabular}{c|cccc}
    %        \toprule[1pt]
    %        X & CrI$_3$ & Li$_{0.125}$CrI$_3$ & Li$_{0.25}$CrI$_3$ & Li$_{0.5}$CrI$_3$ \\
    %        \midrule[1pt]
    %        Valence & \shortstack{100\% Cr$^{3+}$} & \shortstack{75\% Cr$^{3+}$ \\ 25\% Cr$^{2+}$} & \shortstack{50\% Cr$^{3+}$ \\ 50\% Cr$^{2+}$} & \shortstack{100\% Cr$^{2+}$} \\
    %        $J_1$ & -3.162 & -3.486 & -8.667 & -6.694 \\
    %        $J^{super}_1$ & 6.807 & 8.593 & 13.415 & 12.269 \\
    %        $J^{double}_1$ & 0 & -1.707 & -1.947 & 0 \\
    %        \bottomrule[1pt]
    %    \end{tabular}
    %\end{table*}

    \begin{table*}[!htbp]
        \centering
        \caption{The values of $|V_{p d_i}|^2$, and $|E_{p'} - E_{d_j}|$ between $p$ orbitals of I atoms and $d_{z^2}$, $d_{xz}$ 
        orbitals of Cr atoms are presented for CrI$_3$, Li$_{0.125}$CrI$_3$, Li$_{0.25}$CrI$_3$, and Li$_{0.5}$CrI$_3$. The calculation of 
        $J^{super}_1 A$ includes all possible paths connecting two nearest Cr atoms. The proportions of valence change for these materials are 
        listed in the first column. $|V_{p d_i}|$, and $|E_{p'} - E_{d_j}|$ are in units of eV and $J^{super}_1A$ is in 
        units of eV$^3$. }
        \vspace{3pt}
        \setlength{\tabcolsep}{0.45cm}
        \renewcommand{\arraystretch}{1.2}
        \label{table2}
        \begin{tabular}{c|ccccccc}\toprule[1pt]
            %& &\multicolumn{6}{c}{CrI$_3$} \\
            %\cline{3-8}
            CrI$_3$ & & $p_z - d_{z^2}$ & $p_x - d_{z^2}$ & $p_y - d_{z^2}$ & $p_z - d_{xz}$ & $p_x - d_{xz}$ & $p_y - d_{xz}$ \\
            \hline
            \multirow{3}{*}{\shortstack{100\% Cr$^{3+}$}} & $|V_{p d_i}|^2$ & 0.004 & 0.654 & 0.003 & 0.760 & 0.001 & 1.2E-4 \\
            %& $J_{p' d_j}$ & 4.304 & 0.034 & 0.009 & 0.011 & 0.766 & 5.6E-6 \\ 
            & $|E_{p'} - E_{d_j}|$ & 1.172 & 1.176 & 1.173 & 2.016 & 2.013 & 2.012 \\
            \cline{3-8}
            & $J^{super}_1 A$ & \multicolumn{6}{c}{6.807} \\
            \midrule[1pt]
            %& &\multicolumn{6}{c}{Li$_{0.125}$CrI$_3$} \\
            %\cline{3-8}
            Li$_{0.125}$CrI$_3$ & & $p_z - d_{z^2}$ & $p_x - d_{z^2}$ & $p_y - d_{z^2}$ & $p_z - d_{xz}$ & $p_x - d_{xz}$ & $p_y - d_{xz}$ \\
            \hline
            \multirow{3}{*}{\shortstack{75\% Cr$^{3+}$ \\ 25\% Cr$^{2+}$}} & $|V_{p d_i}|^2$ & 0.001 & 0.578 & 0.013 & 0.830 & 0.001 & 2.5E-4 \\
            %& $J_{p' d_j}$ & 5.709 & 0.016 & 0.434 & 0.003 & 1.724 & 0.001 \\
            & $|E_{p'} - E_{d_j}|$ & 0.748 & 0.748 & 0.761 & 0.962 & 0.962 & 0.949 \\
            \cline{3-8}
            & $J^{super}_1 A$ & \multicolumn{6}{c}{8.593} \\
            \midrule[1pt]
            %& &\multicolumn{6}{c}{Li$_{0.25}$CrI$_3$} \\
            %\cline{3-8}
            Li$_{0.25}$CrI$_3$ & & $p_z - d_{z^2}$ & $p_x - d_{z^2}$ & $p_y - d_{z^2}$ & $p_z - d_{xz}$ & $p_x - d_{xz}$ & $p_y - d_{xz}$ \\
            \hline
            \multirow{3}{*}{\shortstack{50\% Cr$^{3+}$ \\ 50\% Cr$^{2+}$}} & $|V_{p d_i}|^2$ & 0.011 & 0.550 & 0.010 & 0.578 & 0.019 & 8.6E-4 \\
            %& $J_{p' d_j}$ & 6.802 & 0.101 & 0.321 & 0.040 & 1.242 & 0.002 \\ 
            & $|E_{p'} - E_{d_j}|$ & 0.586 & 0.585 & 0.634 & 0.929 & 0.930 & 0.882 \\
            \cline{3-8}
            & $J^{super}_1 A$ & \multicolumn{6}{c}{13.415} \\
            \midrule[1pt]
            %& &\multicolumn{6}{c}{Li$_{0.5}$CrI$_3$} \\
            %\cline{3-8}
            Li$_{0.5}$CrI$_3$ & & $p_z - d_{z^2}$ & $p_x - d_{z^2}$ & $p_y - d_{z^2}$ & $p_z - d_{xz}$ & $p_x - d_{xz}$ & $p_y - d_{xz}$ \\
            \hline
            \multirow{3}{*}{\shortstack{100\% Cr$^{2+}$}} & $|V_{p d_i}|^2$ & 0.007 & 0.543 & 0.009 & 0.639 & 0.012 & 7.5E-4 \\
            %& $J_{p' d_j}$ & 7.429 & 0.076 & 0.351 & 0.025 & 1.367 & 0.002 \\
            & $|E_{p'} - E_{d_j}|$ & 0.556 & 0.556 & 0.613 & 0.934 & 0.935 & 0.879 \\
            \cline{3-8}
            & $J^{super}_1 A$ & \multicolumn{6}{c}{12.269} \\
            \bottomrule[1pt]
        \end{tabular}
    \end{table*}

    \par {\it \textcolor{blue}{Superexchange contribution of $J_1$}.}---Synergistic superexchange 
    \cite{anderson1959,goodenough1955,kanamori1960} and double-exchange \cite{zener1951} interaction contribute dominantly 
    to $J_1$ enhancement, as supported by $|E_p-E_d|$ reduction and valence mixing. Superexchange interaction involves two 
    processes (shown in Fig. \ref{Fig3}(a)): direct exchange between Cr$_j$ $d_j$ electrons and I $p'$ electrons 
    (denoted by $J_{p' d_j}$) and hopping process between I $p$ electrons and Cr$_i$ $d_i$ electrons (represented by $|V_{p d_i}|^2/A$). 
    The coupling $J^{super}_1$ between the nearest Cr atoms is expressed as \cite{lisuperexchange2023,yang2025} 
    
    \begin{subequations}
        \begin{equation}
            J^{super}_1 = \frac{1}{4A} \sum_{d_i,p,p',d_j} |V_{p d_i}|^2 J_{p' d_j},
            \label{equ2a}
        \end{equation}
        \begin{equation}
            J_{p' d_j} = \frac{2 |V_{p'd_j}|^2}{|E_{p'} - E_{d_j}|}, 
            \label{equ2b}
        \end{equation}
    \end{subequations}

    \noindent where $A \equiv 1/[1/(E_{d_i d_i'}^{\uparrow\uparrow})^2 - 1/(E_{d_i d_i'}^{\uparrow\downarrow})^2]$ is treated as 
    a pending parameter and assumed to remain unchanged during the intercalation process. $E_{d_i d_i'}^{\uparrow\uparrow}$ and 
    $E_{d_i d_i'}^{\uparrow\downarrow}$ are energies of two $d$ electrons on the same Cr atom with parallel and antiparallel 
    spins, respectively. $V_{p d_i}$ is the hopping parameter between $p$ and $d_i$ electrons. The direct p-d exchange, 
    shown in \ref{equ2b}, is derived from the s-d exchange model following the Schrieffer-Wolff transformation 
    \cite{schrieffer1966}. $E_{p'}$ and $E_{d_j}$ are energy levels of $p'$ and $d_j$ electrons, respectively. 
    All these parameters were obtained using DFT calculations and WANNIER90 \cite{Wannier902008,Wannier902014} code. 

    \par Due to the crystal field, $d$-orbitals of Cr in CrI$_3$ are splitted into two groups: $e_g$ and $t_{2g}$. For 
    Li$_n$CrI$_3$ (n = 0.125, 0.25, 0.5), due to Li intercalation, the degeneracy of the $e_g$ orbitals is lifted, yielding 
    distinct $d_{z^2}$ and $d_{x^2-y^2}$ orbitals. Similarly, the $t_{2g}$ orbitals further split, with $d_{xz}$ becoming 
    distinct from the degenerate $d_{yz}$ and $d_{xy}$ orbitals. 
    %the intercalation of Li atoms contribute the electrons doping and the distortion of the Cr-I octaheral units, 
    %which lead to the enhancement of magnetism and the degeneracy lifting of $e_g$ orbitals of 12.5\%, 50\% and 100\% of Cr atoms, respectively. 
    These changes in orbitals are consistent with the Jahn-Teller (JT) effect \cite{lyu2022,petrov2021}. 
    %The JT effect is a structural distortion in a nonlinear polyatomic system (molecule or solid) associated with a certain 
    %electronic configuration, which can alter the band structures and magnetism. 
    In Li$_n$CrI$_3$, the JT effect is caused by electron doping from Li atoms. WANNIER90 \cite{Wannier902008,Wannier902014} 
    calculations indicate that the main contribution to the increase of $J^{super}_1$ arises from $d_{z^2}$ and $d_{xz}$ 
    orbitals. The values of $|V_{p d_i}|^2$, and $E_{p'} - E_{d_j}$ for a representative Cr-I-Cr path in CrI$_3$ and Li$_n$CrI$_3$ 
    are listed in Table. \ref{table2}. To better demonstrate the mechanism of $J^{super}_1$ enhancement, we plotted the data from 
    Table. \ref{table2} in Fig. \ref{Fig3}. As shown in Fig. \ref{Fig3} (c) and (d), the hopping parameters show no significant 
    change, but the energy difference between $p$ and $d$ electrons decreases dramatically after intercalation. This decrease 
    in $|E_{p'} - E_{d_j}|$ contributes to the enhancement of $J^{super}_1$. 
    %Comparison of the CrI$_3$ and Li$_n$CrI$_3$ results reveals that 
    %the enhancement of the superexchange interaction $J^{super}_1$ is primarily due to the reduced energy difference $|E_{p'} - E_{d_j}|$. 
    %The degree of reduction in energy 
    %difference varies among intercalation systems with different Li nominal stoichiometries, which leads to differences in $J^{super}_1$. 
    The calculation of $J^{super}_1 A$ includes all possible paths connecting two nearest Cr atoms, and the results are listed in Table. \ref{table2}. 
    Fig. \ref{Fig3} (e) demonstrates the ratios of $J_1$ between Li-intercalated CrI$_3$ and non-intercalated CrI$_3$ for both the DFT results 
    ($J_1$ (X)/$J_1$ (CrI$_3$)) and the superexchange model ($J^{super}_1$ (X)/$J^{super}_1$ (CrI$_3$)). Here, X represents different meaterials. 
    The ratio of DFT-calculated $J_1$ between Li$_{0.25}$CrI$_3$ and CrI$_3$ is 2.7, which is larger than the corresponding result of 1.97 
    from the superexchange model. 
    %The superexchange model results for Li$_n$CrI$_3$ (n = 0.125, 0.25, 0.5) are qualitatively consistent with the DFT results. 

    \begin{table}[!htbp]
        \centering
        \caption{Values of parameters $|V_{p d_i}|$, $|V_{d_j p}|$, $E_{d_i} - E_{p}$, and calculated $J^{double}_1$ for 
        Li$_{0.125}$CrI$_3$ and Li$_{0.25}$CrI$_3$. $|V_{p d_i}|$, $V_{d_j p}$, and $E_{d_i} - E_{p}$ are all in units of 
        eV while $J^{double}_1$ is in meV. }
        \vspace{3pt}
        \label{table3}
        \setlength{\tabcolsep}{0.3cm}
        \renewcommand{\arraystretch}{1.2}
        \begin{tabular}{cccc}
            \toprule[1pt]
            Li$_{0.125}$CrI$_3$ & $p_z - d_{z^2}$ & $p_x - d_{z^2}$ & $p_y - d_{z^2}$ \\
            \hline
            $|V_{p d_i}|$ & 0.110 & 0.082 & 0.469 \\
            $|V_{d_j p}|$ & 0.005 & 0.412 & 0.013 \\
            $E_{d_i} - E_{p}$ & 0.896 & 0.897 & 0.909 \\
            \cline{2-4}
            $J^{double}_1$ & \multicolumn{3}{c}{-1.707} \\
            \midrule[1pt]
            Li$_{0.25}$CrI$_3$ & $p_z - d_{z^2}$ & $p_x - d_{z^2}$ & $p_y - d_{z^2}$ \\
            \hline
            $|V_{p d_i}|$ & 0.107 & 0.097 & 0.478 \\
            $|V_{d_j p}|$ & 0.014 & 0.498 & 0.005 \\
            $E_{d_i} - E_{p}$ & 0.586 & 0.585 & 0.634 \\
            \cline{2-4}
            $J^{double}_1$ & \multicolumn{3}{c}{-1.947} \\
            \bottomrule[1pt]
        \end{tabular}
    \end{table}

    \par {\it \textcolor{blue}{Double-exchange contribution of $J_1$}.}---The discrepancy between the DFT and superexchange model results 
    may be attributed to double-exchange interaction. The double-exchange interaction (shown in Fig. \ref{Fig3}(b)) explains ferromagnetism 
    in materials with mixed-valence transition metal ions. 
    %It arises from the interplay between electron hopping and Hund's coupling, enabling efficient electron transfer between ions with different valence states. 
    For Li$_n$CrI$_3$ (n = 0.125, 0.25), electron doping from Li atoms alters the valence state of Cr, leading to valence mixing, 
    as shown in Table. S3. Experimentally, valence mixing is also observed, as a significant proportion of Cr$^{3+}$ is reduced to 
    Cr$^{2+}$ in Li$_{0.149}$CrI$_3$ crystals, as confirmed by X-ray photoelectron spectrometry (XPS) \cite{wang2024}. The 
    double-exchange coupling $J^{double}_1$ can be expressed as \cite{SMs} 

    \begin{equation}
        J^{double}_1 = -\frac{1}{4S^2} \sum_{d_i,d_j = d_{z^2}} \sum_{p} |\frac{V_{p d_i} V_{d_j p}}{E_{d_i} - E_p - 3 J_H + 3 U'}|, 
        \label{equ3}
    \end{equation}

    \noindent where $E_{d_i}$ denotes the energy level of $d_{z^2}$ orbital. $J_H$ is Hund coupling between $e_g$ and $t_{2g}$ orbitals. 
    The Coulomb parameter $U'$ corresponds to the interaction between orbitals on a given site, with a relationship $U' = U - 2 J_H$ 
    \cite{maekawa2004}. $U$ is the interaction within orbitals on a given site. $S$ is the size of local spin. The values of 
    $V_{p d_i}$, $V_{d_j p}$, and $E_{d_i} - E_{p}$ for a representative path in Li$_{0.125}$CrI$_3$ and Li$_{0.25}$CrI$_3$ are listed 
    in Table. \ref{table3}. $J^{double}_1$ was calculated by including all possible paths. The ratios $J^{double}_1$ (X)/$J^{double}_1$ 
    (Li$_{0.125}$CrI$_3$) %where X represents Li$_{0.125}$CrI$_3$ and Li$_{0.25}$CrI$_3$, 
    are also shown in Fig. \ref{Fig3} (c) by a red line. The enhancement of $J^{double}_1$ mainly comes from the reduction in 
    energy difference $|E_{p'} - E_{d_j}|$. The superexchange and double-exchange model results for intercalated materials are 
    qualitatively consistent with the DFT calculations.  
    %Both the enhancement of $J^{super}_1$ and $J^{double}_1$ contribute to the enhancement of $J_1$ and $T{\rm{_C}}$. 

    \par The $J_2$ and $J_1'$ couplings are mediated by Cr-I-I-Cr hybridization, a mechanism known as super-super-exchange 
    (SSE) interaction \cite{sivadas2018}. Increasing Li intercalation from 0.125 to 0.25 boosts the $T{\rm{_C}}$ by 
    strengthening both $J_1$ and $J_2$. However, further increasing Li to 0.5 causes $T{\rm{_C}}$ to 
    plateau because the dramatic increase in $J_1'$ is offset by a reduction in $J_1$. 

    %The enhancement of $J_1'$ can be qualitatively understood through increased direct exchange between Cr and I atoms and strengthened 
    %interlayer I-I hybridization. The latter, quantified by $|V_{pp}|^2$ obtained from WANNIER90 calculations, exhibits an obvious 
    %enhancement (values are listed in the SM \cite{SMs}). 

    \begin{figure}[tphb]
    	\centering
        \includegraphics[scale=0.32]{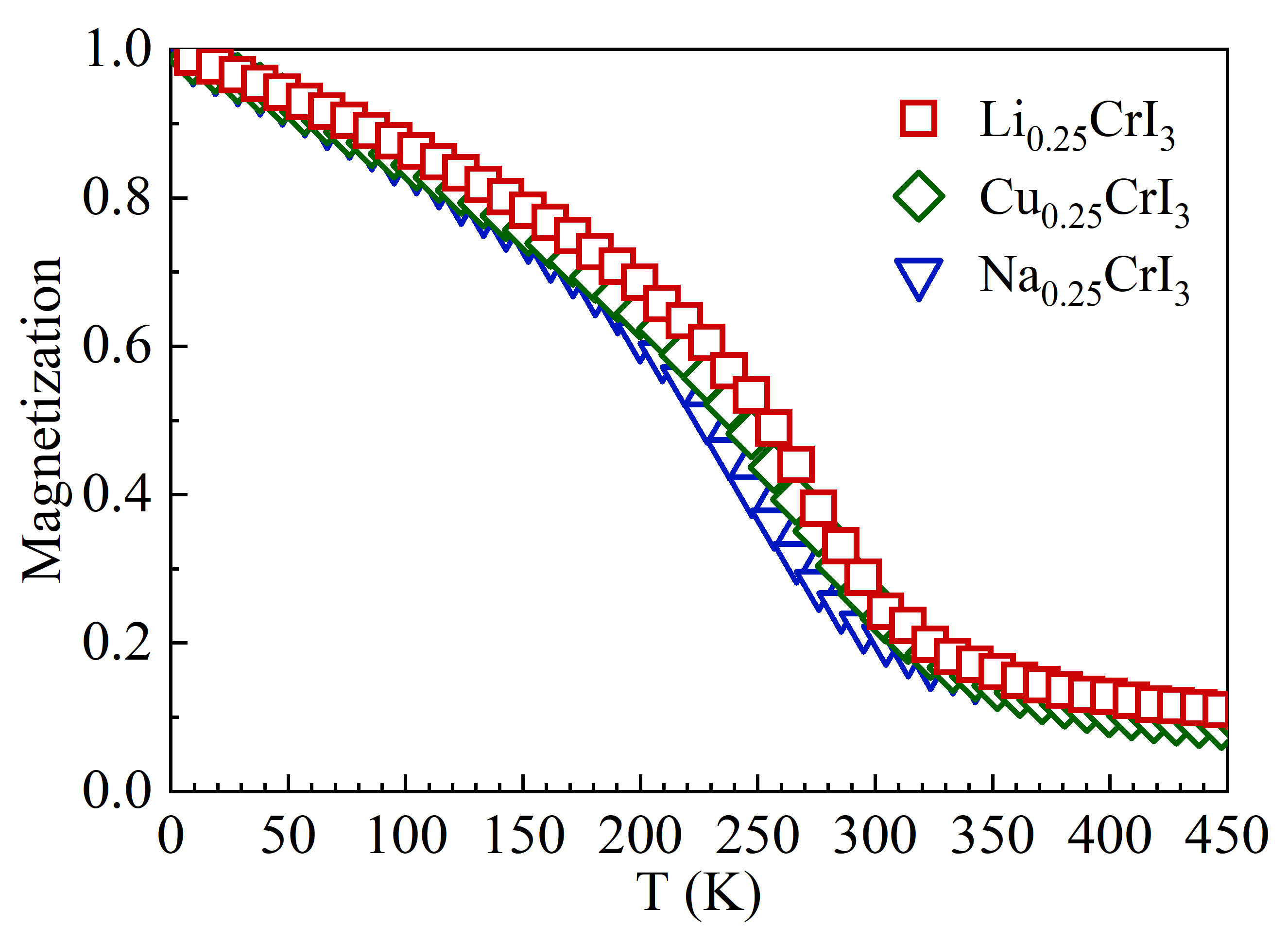}
        \caption{
            Comparison of magnetization curves for Li$_{0.25}$CrI$_3$, Cu$_{0.25}$CrI$_3$, and Na$_{0.25}$CrI$_3$ 
            obtained from MC simulations. 
    	}\label{Fig5}
    \end{figure}

    \par {\it \textcolor{blue}{Prediction of new intercalation materials.}}---We computationally screened seven experimentally 
    accessible intercalation elements (Cu, Na, K, Mg, Ca, Al, Pb) - many previously intercalated in layered materials 
    \cite{huempfner2022,huempfner2023,kanetani2012,kim2024,liK2024,liuCu2017,nongCu2025,scholzel2024,zhangK2020,yaoCu2014,
    zhaoCu2024,tonti2000,toyama2022,wangNaAl2015,weber2019,naik2021} - at a fixed stoichiometry of 0.25. As 
    summarized in Table. S4, Cu$_{0.25}$CrI$_3$ and Na$_{0.25}$CrI$_3$ exhibit $T{\rm{_C}}$ of 267 K and 247 K 
    respectively. %Monte Carlo simulations reveal their distinct magnetization profiles. 
    The enhanced $T{\rm{_C}}$ in Cu$_{0.25}$CrI$_3$ originates from simultaneous strengthening of intralayer 
    couplings ($J_1$ $\uparrow$ 220\%, $J_2$ $\uparrow$ 1257\%) and interlayer AFM $\to$ FM transition 
    without significant $J_1'$ modification. Superexchange and double-exchange modeling (Fig. S6) confirms that $e_g$/$t_{2g}$ 
    orbital splitting and reduced $|E_p-E_d|$ - identical to Li-intercalation mechanisms - drive this enhancement. 
    This transferable response validates intercalation as a general $T{\rm{_C}}$ - engineering strategy for 2D magnets.

    \par {\it \textcolor{blue}{Conclusion.}}---This work establishes a broadly applicable intercalation-driven 
    mechanism for enhancing $T{\rm{_C}}$ in 2D vdW magnets, primarily through reduction of the $|E_p-E_d|$ 
    energy difference between I-p and Cr-d orbitals. This reduction 
    amplifies both superexchange and double-exchange interactions, as evidenced in Li-intercalated CrI$_3$, where 
    first-principles calculations predict a $T{\rm{_C}}$ increase from 57 K to 286 K - qualitatively consistent with 
    experimental reports of up to 420 K, highlighting the synergy between theoretical modeling and experimental 
    realization for Li-intercalation. Extending this mechanism, we identify two novel near 
    RT ferromagnetic semiconductors - Cu$_{0.25}$CrI$_3$ and Na$_{0.25}$CrI$_3$. 
    These findings bridge fundamental theory with practical applications, 
    delivering: (1) A design principle for $T{\rm{_C}}$ enhancement via targeted orbital energy modulation; 
    (2) Experimentally accessible materials for energy-efficient spintronic devices; (3) Systematic validation 
    of intercalation as a scalable strategy for engineering high-$T{\rm{_C}}$ magnetic semiconductors. While the 
    strategy shows transferable efficacy in Cu/Na systems, further experimental synthesis is needed to confirm 
    its full scope.

    \section*{Acknowledgements}
    This work is supported by National Key R\&D Program of China (Grant No. 2022YFA1405100), Chinese Academy of Sciences Project for 
    Young Scientists in Basic Research (Grant No. YSBR-030), and Basic Research Program of the Chinese Academy of Sciences Based on 
    Major Scientific Infrastructures (Grant No. JZHKYPT-2021-08). GS was supported in part by the Innovation Program for Quantum 
    Science and Technology under Grant No. 2024ZD0300500, NSFC No. 12447101 and the Strategic Priority Research Program of Chinese 
    Academy of Sciences (Grant No. XDB1270000). 

    %\newpage
    %\bibliographystyle{apsrev4-2}
    %\bibliography{refCrI3}

\begin{thebibliography}{69}%
\makeatletter
\providecommand \@ifxundefined [1]{%
 \@ifx{#1\undefined}
}%
\providecommand \@ifnum [1]{%
 \ifnum #1\expandafter \@firstoftwo
 \else \expandafter \@secondoftwo
 \fi
}%
\providecommand \@ifx [1]{%
 \ifx #1\expandafter \@firstoftwo
 \else \expandafter \@secondoftwo
 \fi
}%
\providecommand \natexlab [1]{#1}%
\providecommand \enquote  [1]{``#1''}%
\providecommand \bibnamefont  [1]{#1}%
\providecommand \bibfnamefont [1]{#1}%
\providecommand \citenamefont [1]{#1}%
\providecommand \href@noop [0]{\@secondoftwo}%
\providecommand \href [0]{\begingroup \@sanitize@url \@href}%
\providecommand \@href[1]{\@@startlink{#1}\@@href}%
\providecommand \@@href[1]{\endgroup#1\@@endlink}%
\providecommand \@sanitize@url [0]{\catcode `\\12\catcode `\$12\catcode
  `\&12\catcode `\#12\catcode `\^12\catcode `\_12\catcode `\%12\relax}%
\providecommand \@@startlink[1]{}%
\providecommand \@@endlink[0]{}%
\providecommand \url  [0]{\begingroup\@sanitize@url \@url }%
\providecommand \@url [1]{\endgroup\@href {#1}{\urlprefix }}%
\providecommand \urlprefix  [0]{URL }%
\providecommand \Eprint [0]{\href }%
\providecommand \doibase [0]{https://doi.org/}%
\providecommand \selectlanguage [0]{\@gobble}%
\providecommand \bibinfo  [0]{\@secondoftwo}%
\providecommand \bibfield  [0]{\@secondoftwo}%
\providecommand \translation [1]{[#1]}%
\providecommand \BibitemOpen [0]{}%
\providecommand \bibitemStop [0]{}%
\providecommand \bibitemNoStop [0]{.\EOS\space}%
\providecommand \EOS [0]{\spacefactor3000\relax}%
\providecommand \BibitemShut  [1]{\csname bibitem#1\endcsname}%
\let\auto@bib@innerbib\@empty
%</preamble>
\bibitem [{\citenamefont {Huang}\ \emph {et~al.}(2017)\citenamefont {Huang},
  \citenamefont {Clark}, \citenamefont {{Navarro-Moratalla}}, \citenamefont
  {Klein}, \citenamefont {Cheng}, \citenamefont {Seyler}, \citenamefont
  {Zhong}, \citenamefont {Schmidgall}, \citenamefont {McGuire}, \citenamefont
  {Cobden}, \citenamefont {Yao}, \citenamefont {Xiao}, \citenamefont
  {{Jarillo-Herrero}},\ and\ \citenamefont {Xu}}]{huangCrI32017}%
  \BibitemOpen
  \bibfield  {author} {\bibinfo {author} {\bibfnamefont {B.}~\bibnamefont
  {Huang}}, \bibinfo {author} {\bibfnamefont {G.}~\bibnamefont {Clark}},
  \bibinfo {author} {\bibfnamefont {E.}~\bibnamefont {{Navarro-Moratalla}}},
  \bibinfo {author} {\bibfnamefont {D.~R.}\ \bibnamefont {Klein}}, \bibinfo
  {author} {\bibfnamefont {R.}~\bibnamefont {Cheng}}, \bibinfo {author}
  {\bibfnamefont {K.~L.}\ \bibnamefont {Seyler}}, \bibinfo {author}
  {\bibfnamefont {D.}~\bibnamefont {Zhong}}, \bibinfo {author} {\bibfnamefont
  {E.}~\bibnamefont {Schmidgall}}, \bibinfo {author} {\bibfnamefont {M.~A.}\
  \bibnamefont {McGuire}}, \bibinfo {author} {\bibfnamefont {D.~H.}\
  \bibnamefont {Cobden}}, \bibinfo {author} {\bibfnamefont {W.}~\bibnamefont
  {Yao}}, \bibinfo {author} {\bibfnamefont {D.}~\bibnamefont {Xiao}}, \bibinfo
  {author} {\bibfnamefont {P.}~\bibnamefont {{Jarillo-Herrero}}},\ and\
  \bibinfo {author} {\bibfnamefont {X.}~\bibnamefont {Xu}},\ }\bibfield
  {title} {\bibinfo {title} {Layer-dependent ferromagnetism in a van der
  {{Waals}} crystal down to the monolayer limit},\ }\href
  {https://doi.org/10.1038/nature22391} {\bibfield  {journal} {\bibinfo
  {journal} {Nature}\ }\textbf {\bibinfo {volume} {546}},\ \bibinfo {pages}
  {270} (\bibinfo {year} {2017})}\BibitemShut {NoStop}%
\bibitem [{\citenamefont {Lee}\ \emph {et~al.}(2021)\citenamefont {Lee},
  \citenamefont {Dismukes}, \citenamefont {Telford}, \citenamefont {Wiscons},
  \citenamefont {Wang}, \citenamefont {Xu}, \citenamefont {Nuckolls},
  \citenamefont {Dean}, \citenamefont {Roy},\ and\ \citenamefont
  {Zhu}}]{lee2021}%
  \BibitemOpen
  \bibfield  {author} {\bibinfo {author} {\bibfnamefont {K.}~\bibnamefont
  {Lee}}, \bibinfo {author} {\bibfnamefont {A.~H.}\ \bibnamefont {Dismukes}},
  \bibinfo {author} {\bibfnamefont {E.~J.}\ \bibnamefont {Telford}}, \bibinfo
  {author} {\bibfnamefont {R.~A.}\ \bibnamefont {Wiscons}}, \bibinfo {author}
  {\bibfnamefont {J.}~\bibnamefont {Wang}}, \bibinfo {author} {\bibfnamefont
  {X.}~\bibnamefont {Xu}}, \bibinfo {author} {\bibfnamefont {C.}~\bibnamefont
  {Nuckolls}}, \bibinfo {author} {\bibfnamefont {C.~R.}\ \bibnamefont {Dean}},
  \bibinfo {author} {\bibfnamefont {X.}~\bibnamefont {Roy}},\ and\ \bibinfo
  {author} {\bibfnamefont {X.}~\bibnamefont {Zhu}},\ }\bibfield  {title}
  {\bibinfo {title} {Magnetic {{Order}} and {{Symmetry}} in the {{2D
  Semiconductor CrSBr}}},\ }\href
  {https://doi.org/10.1021/acs.nanolett.1c00219} {\bibfield  {journal}
  {\bibinfo  {journal} {Nano Lett.}\ }\textbf {\bibinfo {volume} {21}},\
  \bibinfo {pages} {3511} (\bibinfo {year} {2021})}\BibitemShut {NoStop}%
\bibitem [{\citenamefont {Gong}\ \emph {et~al.}(2017)\citenamefont {Gong},
  \citenamefont {Li}, \citenamefont {Li}, \citenamefont {Ji}, \citenamefont
  {Stern}, \citenamefont {Xia}, \citenamefont {Cao}, \citenamefont {Bao},
  \citenamefont {Wang}, \citenamefont {Wang}, \citenamefont {Qiu},
  \citenamefont {Cava}, \citenamefont {Louie}, \citenamefont {Xia},\ and\
  \citenamefont {Zhang}}]{gong2017}%
  \BibitemOpen
  \bibfield  {author} {\bibinfo {author} {\bibfnamefont {C.}~\bibnamefont
  {Gong}}, \bibinfo {author} {\bibfnamefont {L.}~\bibnamefont {Li}}, \bibinfo
  {author} {\bibfnamefont {Z.}~\bibnamefont {Li}}, \bibinfo {author}
  {\bibfnamefont {H.}~\bibnamefont {Ji}}, \bibinfo {author} {\bibfnamefont
  {A.}~\bibnamefont {Stern}}, \bibinfo {author} {\bibfnamefont
  {Y.}~\bibnamefont {Xia}}, \bibinfo {author} {\bibfnamefont {T.}~\bibnamefont
  {Cao}}, \bibinfo {author} {\bibfnamefont {W.}~\bibnamefont {Bao}}, \bibinfo
  {author} {\bibfnamefont {C.}~\bibnamefont {Wang}}, \bibinfo {author}
  {\bibfnamefont {Y.}~\bibnamefont {Wang}}, \bibinfo {author} {\bibfnamefont
  {Z.~Q.}\ \bibnamefont {Qiu}}, \bibinfo {author} {\bibfnamefont {R.~J.}\
  \bibnamefont {Cava}}, \bibinfo {author} {\bibfnamefont {S.~G.}\ \bibnamefont
  {Louie}}, \bibinfo {author} {\bibfnamefont {J.}~\bibnamefont {Xia}},\ and\
  \bibinfo {author} {\bibfnamefont {X.}~\bibnamefont {Zhang}},\ }\bibfield
  {title} {\bibinfo {title} {Discovery of intrinsic ferromagnetism in
  two-dimensional van der {{Waals}} crystals},\ }\href
  {https://doi.org/10.1038/nature22060} {\bibfield  {journal} {\bibinfo
  {journal} {Nature}\ }\textbf {\bibinfo {volume} {546}},\ \bibinfo {pages}
  {265} (\bibinfo {year} {2017})}\BibitemShut {NoStop}%
\bibitem [{\citenamefont {Deng}\ \emph {et~al.}(2018)\citenamefont {Deng},
  \citenamefont {Yu}, \citenamefont {Song}, \citenamefont {Zhang},
  \citenamefont {Wang}, \citenamefont {Sun}, \citenamefont {Yi}, \citenamefont
  {Wu}, \citenamefont {Wu}, \citenamefont {Zhu}, \citenamefont {Wang},
  \citenamefont {Chen},\ and\ \citenamefont {Zhang}}]{dengGate2018}%
  \BibitemOpen
  \bibfield  {author} {\bibinfo {author} {\bibfnamefont {Y.}~\bibnamefont
  {Deng}}, \bibinfo {author} {\bibfnamefont {Y.}~\bibnamefont {Yu}}, \bibinfo
  {author} {\bibfnamefont {Y.}~\bibnamefont {Song}}, \bibinfo {author}
  {\bibfnamefont {J.}~\bibnamefont {Zhang}}, \bibinfo {author} {\bibfnamefont
  {N.~Z.}\ \bibnamefont {Wang}}, \bibinfo {author} {\bibfnamefont
  {Z.}~\bibnamefont {Sun}}, \bibinfo {author} {\bibfnamefont {Y.}~\bibnamefont
  {Yi}}, \bibinfo {author} {\bibfnamefont {Y.~Z.}\ \bibnamefont {Wu}}, \bibinfo
  {author} {\bibfnamefont {S.}~\bibnamefont {Wu}}, \bibinfo {author}
  {\bibfnamefont {J.}~\bibnamefont {Zhu}}, \bibinfo {author} {\bibfnamefont
  {J.}~\bibnamefont {Wang}}, \bibinfo {author} {\bibfnamefont {X.~H.}\
  \bibnamefont {Chen}},\ and\ \bibinfo {author} {\bibfnamefont
  {Y.}~\bibnamefont {Zhang}},\ }\bibfield  {title} {\bibinfo {title}
  {Gate-tunable room-temperature ferromagnetism in two-dimensional
  {{Fe{\textsubscript{3}}GeTe{\textsubscript{2}}}}},\ }\href
  {https://doi.org/10.1038/s41586-018-0626-9} {\bibfield  {journal} {\bibinfo
  {journal} {Nature}\ }\textbf {\bibinfo {volume} {563}},\ \bibinfo {pages}
  {94} (\bibinfo {year} {2018})}\BibitemShut {NoStop}%
\bibitem [{\citenamefont {Jiang}\ \emph
  {et~al.}(2018{\natexlab{a}})\citenamefont {Jiang}, \citenamefont {Shan},\
  and\ \citenamefont {Mak}}]{jiangElectric2018}%
  \BibitemOpen
  \bibfield  {author} {\bibinfo {author} {\bibfnamefont {S.}~\bibnamefont
  {Jiang}}, \bibinfo {author} {\bibfnamefont {J.}~\bibnamefont {Shan}},\ and\
  \bibinfo {author} {\bibfnamefont {K.~F.}\ \bibnamefont {Mak}},\ }\bibfield
  {title} {\bibinfo {title} {Electric-field switching of two-dimensional van
  der {{Waals}} magnets},\ }\href {https://doi.org/10.1038/s41563-018-0040-6}
  {\bibfield  {journal} {\bibinfo  {journal} {Nat. Mater.}\ }\textbf {\bibinfo
  {volume} {17}},\ \bibinfo {pages} {406} (\bibinfo {year}
  {2018}{\natexlab{a}})}\BibitemShut {NoStop}%
\bibitem [{\citenamefont {{El-Habib}}\ \emph {et~al.}(2024)\citenamefont
  {{El-Habib}}, \citenamefont {Brioual}, \citenamefont {Bouachri},
  \citenamefont {Zimou}, \citenamefont {Aouni}, \citenamefont {Diani},\ and\
  \citenamefont {Addou}}]{el-habib2024}%
  \BibitemOpen
  \bibfield  {author} {\bibinfo {author} {\bibfnamefont {A.}~\bibnamefont
  {{El-Habib}}}, \bibinfo {author} {\bibfnamefont {B.}~\bibnamefont {Brioual}},
  \bibinfo {author} {\bibfnamefont {M.}~\bibnamefont {Bouachri}}, \bibinfo
  {author} {\bibfnamefont {J.}~\bibnamefont {Zimou}}, \bibinfo {author}
  {\bibfnamefont {A.}~\bibnamefont {Aouni}}, \bibinfo {author} {\bibfnamefont
  {M.}~\bibnamefont {Diani}},\ and\ \bibinfo {author} {\bibfnamefont
  {M.}~\bibnamefont {Addou}},\ }\bibfield  {title} {\bibinfo {title} {Synthesis
  and characterization of {{Nd-doped CeO{\textsubscript{2}}}} thin films grown
  by spray pyrolysis method: {{Structural}}, optical and electrochemical
  properties},\ }\href {https://doi.org/10.1016/j.surfin.2024.103859}
  {\bibfield  {journal} {\bibinfo  {journal} {Surf. Interf.}\ }\textbf
  {\bibinfo {volume} {45}},\ \bibinfo {pages} {103859} (\bibinfo {year}
  {2024})}\BibitemShut {NoStop}%
\bibitem [{\citenamefont {Jiang}\ \emph
  {et~al.}(2018{\natexlab{b}})\citenamefont {Jiang}, \citenamefont {Li},
  \citenamefont {Wang}, \citenamefont {Mak},\ and\ \citenamefont
  {Shan}}]{jiangdoping2018}%
  \BibitemOpen
  \bibfield  {author} {\bibinfo {author} {\bibfnamefont {S.}~\bibnamefont
  {Jiang}}, \bibinfo {author} {\bibfnamefont {L.}~\bibnamefont {Li}}, \bibinfo
  {author} {\bibfnamefont {Z.}~\bibnamefont {Wang}}, \bibinfo {author}
  {\bibfnamefont {K.~F.}\ \bibnamefont {Mak}},\ and\ \bibinfo {author}
  {\bibfnamefont {J.}~\bibnamefont {Shan}},\ }\bibfield  {title} {\bibinfo
  {title} {Controlling magnetism in {{2D CrI{\textsubscript{3}}}} by
  electrostatic doping},\ }\href {https://doi.org/10.1038/s41565-018-0135-x}
  {\bibfield  {journal} {\bibinfo  {journal} {Nat. Nanotechnol.}\ }\textbf
  {\bibinfo {volume} {13}},\ \bibinfo {pages} {549} (\bibinfo {year}
  {2018}{\natexlab{b}})}\BibitemShut {NoStop}%
\bibitem [{\citenamefont {Ni}\ \emph {et~al.}(2021)\citenamefont {Ni},
  \citenamefont {Haglund}, \citenamefont {Wang}, \citenamefont {Xu},
  \citenamefont {Bernhard}, \citenamefont {Mandrus}, \citenamefont {Qian},
  \citenamefont {Mele}, \citenamefont {Kane},\ and\ \citenamefont
  {Wu}}]{nistrain2021}%
  \BibitemOpen
  \bibfield  {author} {\bibinfo {author} {\bibfnamefont {Z.}~\bibnamefont
  {Ni}}, \bibinfo {author} {\bibfnamefont {A.~V.}\ \bibnamefont {Haglund}},
  \bibinfo {author} {\bibfnamefont {H.}~\bibnamefont {Wang}}, \bibinfo {author}
  {\bibfnamefont {B.}~\bibnamefont {Xu}}, \bibinfo {author} {\bibfnamefont
  {C.}~\bibnamefont {Bernhard}}, \bibinfo {author} {\bibfnamefont {D.~G.}\
  \bibnamefont {Mandrus}}, \bibinfo {author} {\bibfnamefont {X.}~\bibnamefont
  {Qian}}, \bibinfo {author} {\bibfnamefont {E.~J.}\ \bibnamefont {Mele}},
  \bibinfo {author} {\bibfnamefont {C.~L.}\ \bibnamefont {Kane}},\ and\
  \bibinfo {author} {\bibfnamefont {L.}~\bibnamefont {Wu}},\ }\bibfield
  {title} {\bibinfo {title} {Imaging the {{N{\'e}el}} vector switching in the
  monolayer antiferromagnet {{MnPSe{\textsubscript{3}}}} with strain-controlled
  {{Ising}} order},\ }\href {https://doi.org/10.1038/s41565-021-00885-5}
  {\bibfield  {journal} {\bibinfo  {journal} {Nat. Nanotechnol.}\ }\textbf
  {\bibinfo {volume} {16}},\ \bibinfo {pages} {782} (\bibinfo {year}
  {2021})}\BibitemShut {NoStop}%
\bibitem [{\citenamefont {O'Neill}\ \emph {et~al.}(2023)\citenamefont
  {O'Neill}, \citenamefont {Rahman}, \citenamefont {Zhang}, \citenamefont
  {Schoenherr}, \citenamefont {Yildirim}, \citenamefont {Gu}, \citenamefont
  {Su}, \citenamefont {Lu},\ and\ \citenamefont {Seidel}}]{oneill2023}%
  \BibitemOpen
  \bibfield  {author} {\bibinfo {author} {\bibfnamefont {A.}~\bibnamefont
  {O'Neill}}, \bibinfo {author} {\bibfnamefont {S.}~\bibnamefont {Rahman}},
  \bibinfo {author} {\bibfnamefont {Z.}~\bibnamefont {Zhang}}, \bibinfo
  {author} {\bibfnamefont {P.}~\bibnamefont {Schoenherr}}, \bibinfo {author}
  {\bibfnamefont {T.}~\bibnamefont {Yildirim}}, \bibinfo {author}
  {\bibfnamefont {B.}~\bibnamefont {Gu}}, \bibinfo {author} {\bibfnamefont
  {G.}~\bibnamefont {Su}}, \bibinfo {author} {\bibfnamefont {Y.}~\bibnamefont
  {Lu}},\ and\ \bibinfo {author} {\bibfnamefont {J.}~\bibnamefont {Seidel}},\
  }\bibfield  {title} {\bibinfo {title} {Enhanced {{Room Temperature
  Ferromagnetism}} in {{Highly Strained 2D Semiconductor
  Cr}}{\textsubscript{2}}{{Ge}}{\textsubscript{2}}{{Te}}{\textsubscript{6}}},\
  }\href {https://doi.org/10.1021/acsnano.2c10209} {\bibfield  {journal}
  {\bibinfo  {journal} {ACS Nano}\ }\textbf {\bibinfo {volume} {17}},\ \bibinfo
  {pages} {735} (\bibinfo {year} {2023})}\BibitemShut {NoStop}%
\bibitem [{\citenamefont {Zeng}\ \emph {et~al.}(2024)\citenamefont {Zeng},
  \citenamefont {Yu}, \citenamefont {Liu}, \citenamefont {Lu}, \citenamefont
  {Wei}, \citenamefont {Gao}, \citenamefont {Hong}, \citenamefont {Zhang},
  \citenamefont {Zhang},\ and\ \citenamefont {Zhang}}]{zeng}%
  \BibitemOpen
  \bibfield  {author} {\bibinfo {author} {\bibfnamefont {H.}~\bibnamefont
  {Zeng}}, \bibinfo {author} {\bibfnamefont {H.}~\bibnamefont {Yu}}, \bibinfo
  {author} {\bibfnamefont {B.}~\bibnamefont {Liu}}, \bibinfo {author}
  {\bibfnamefont {S.}~\bibnamefont {Lu}}, \bibinfo {author} {\bibfnamefont
  {X.}~\bibnamefont {Wei}}, \bibinfo {author} {\bibfnamefont {L.}~\bibnamefont
  {Gao}}, \bibinfo {author} {\bibfnamefont {M.}~\bibnamefont {Hong}}, \bibinfo
  {author} {\bibfnamefont {X.}~\bibnamefont {Zhang}}, \bibinfo {author}
  {\bibfnamefont {Z.}~\bibnamefont {Zhang}},\ and\ \bibinfo {author}
  {\bibfnamefont {Y.}~\bibnamefont {Zhang}},\ }\bibfield  {title} {\bibinfo
  {title} {Gradient-{{Strained Van Der Waals Heterojunctions}} for
  {{High-Efficient Photodetectors}}},\ }\href
  {https://doi.org/10.1002/adfm.202400712} {\bibfield  {journal} {\bibinfo
  {journal} {Adv. Funct. Mater.}\ }\textbf {\bibinfo {volume} {34}},\ \bibinfo
  {pages} {2400712} (\bibinfo {year} {2024})}\BibitemShut {NoStop}%
\bibitem [{\citenamefont {You}\ \emph {et~al.}(2023)\citenamefont {You},
  \citenamefont {Dong}, \citenamefont {Gu},\ and\ \citenamefont
  {Su}}]{youP2023}%
  \BibitemOpen
  \bibfield  {author} {\bibinfo {author} {\bibfnamefont {J.-Y.}\ \bibnamefont
  {You}}, \bibinfo {author} {\bibfnamefont {X.-J.}\ \bibnamefont {Dong}},
  \bibinfo {author} {\bibfnamefont {B.}~\bibnamefont {Gu}},\ and\ \bibinfo
  {author} {\bibfnamefont {G.}~\bibnamefont {Su}},\ }\bibfield  {title}
  {\bibinfo {title} {Possible {{Room-Temperature Ferromagnetic
  Semiconductors}}},\ }\href {https://doi.org/10.1088/0256-307X/40/6/067502}
  {\bibfield  {journal} {\bibinfo  {journal} {Chin. Phys. Lett.}\ }\textbf
  {\bibinfo {volume} {40}},\ \bibinfo {pages} {067502} (\bibinfo {year}
  {2023})}\BibitemShut {NoStop}%
\bibitem [{\citenamefont {Datta}\ \emph {et~al.}(2017)\citenamefont {Datta},
  \citenamefont {Cai}, \citenamefont {Yudhistira}, \citenamefont {Zeng},
  \citenamefont {Zhang}, \citenamefont {Zhang}, \citenamefont {Adam},
  \citenamefont {Wu},\ and\ \citenamefont {Loh}}]{datta2017}%
  \BibitemOpen
  \bibfield  {author} {\bibinfo {author} {\bibfnamefont {S.}~\bibnamefont
  {Datta}}, \bibinfo {author} {\bibfnamefont {Y.}~\bibnamefont {Cai}}, \bibinfo
  {author} {\bibfnamefont {I.}~\bibnamefont {Yudhistira}}, \bibinfo {author}
  {\bibfnamefont {Z.}~\bibnamefont {Zeng}}, \bibinfo {author} {\bibfnamefont
  {Y.-W.}\ \bibnamefont {Zhang}}, \bibinfo {author} {\bibfnamefont
  {H.}~\bibnamefont {Zhang}}, \bibinfo {author} {\bibfnamefont
  {S.}~\bibnamefont {Adam}}, \bibinfo {author} {\bibfnamefont {J.}~\bibnamefont
  {Wu}},\ and\ \bibinfo {author} {\bibfnamefont {K.~P.}\ \bibnamefont {Loh}},\
  }\bibfield  {title} {\bibinfo {title} {Tuning magnetoresistance in molybdenum
  disulphide and graphene using a molecular spin transition},\ }\href
  {https://doi.org/10.1038/s41467-017-00727-w} {\bibfield  {journal} {\bibinfo
  {journal} {Nat. Commun.}\ }\textbf {\bibinfo {volume} {8}},\ \bibinfo {pages}
  {677} (\bibinfo {year} {2017})}\BibitemShut {NoStop}%
\bibitem [{\citenamefont {{Gonz{\'a}lez-Herrero}}\ \emph
  {et~al.}(2016)\citenamefont {{Gonz{\'a}lez-Herrero}}, \citenamefont
  {{G{\'o}mez-Rodr{\'i}guez}}, \citenamefont {Mallet}, \citenamefont {Moaied},
  \citenamefont {Palacios}, \citenamefont {Salgado}, \citenamefont {Ugeda},
  \citenamefont {Veuillen}, \citenamefont {Yndurain},\ and\ \citenamefont
  {Brihuega}}]{gonzalez2016}%
  \BibitemOpen
  \bibfield  {author} {\bibinfo {author} {\bibfnamefont {H.}~\bibnamefont
  {{Gonz{\'a}lez-Herrero}}}, \bibinfo {author} {\bibfnamefont {J.~M.}\
  \bibnamefont {{G{\'o}mez-Rodr{\'i}guez}}}, \bibinfo {author} {\bibfnamefont
  {P.}~\bibnamefont {Mallet}}, \bibinfo {author} {\bibfnamefont
  {M.}~\bibnamefont {Moaied}}, \bibinfo {author} {\bibfnamefont {J.~J.}\
  \bibnamefont {Palacios}}, \bibinfo {author} {\bibfnamefont {C.}~\bibnamefont
  {Salgado}}, \bibinfo {author} {\bibfnamefont {M.~M.}\ \bibnamefont {Ugeda}},
  \bibinfo {author} {\bibfnamefont {J.-Y.}\ \bibnamefont {Veuillen}}, \bibinfo
  {author} {\bibfnamefont {F.}~\bibnamefont {Yndurain}},\ and\ \bibinfo
  {author} {\bibfnamefont {I.}~\bibnamefont {Brihuega}},\ }\bibfield  {title}
  {\bibinfo {title} {Atomic-scale control of graphene magnetism by using
  hydrogen atoms},\ }\href {https://doi.org/10.1126/science.aad8038} {\bibfield
   {journal} {\bibinfo  {journal} {Science}\ }\textbf {\bibinfo {volume}
  {352}},\ \bibinfo {pages} {437} (\bibinfo {year} {2016})}\BibitemShut
  {NoStop}%
\bibitem [{\citenamefont {Rajapakse}\ \emph {et~al.}(2021)\citenamefont
  {Rajapakse}, \citenamefont {Karki}, \citenamefont {Abu}, \citenamefont
  {Pishgar}, \citenamefont {Musa}, \citenamefont {Riyadh}, \citenamefont {Yu},
  \citenamefont {Sumanasekera},\ and\ \citenamefont
  {Jasinski}}]{rajapakse2021}%
  \BibitemOpen
  \bibfield  {author} {\bibinfo {author} {\bibfnamefont {M.}~\bibnamefont
  {Rajapakse}}, \bibinfo {author} {\bibfnamefont {B.}~\bibnamefont {Karki}},
  \bibinfo {author} {\bibfnamefont {U.~O.}\ \bibnamefont {Abu}}, \bibinfo
  {author} {\bibfnamefont {S.}~\bibnamefont {Pishgar}}, \bibinfo {author}
  {\bibfnamefont {M.~R.~K.}\ \bibnamefont {Musa}}, \bibinfo {author}
  {\bibfnamefont {S.~M.~S.}\ \bibnamefont {Riyadh}}, \bibinfo {author}
  {\bibfnamefont {M.}~\bibnamefont {Yu}}, \bibinfo {author} {\bibfnamefont
  {G.}~\bibnamefont {Sumanasekera}},\ and\ \bibinfo {author} {\bibfnamefont
  {J.~B.}\ \bibnamefont {Jasinski}},\ }\bibfield  {title} {\bibinfo {title}
  {Intercalation as a versatile tool for fabrication, property tuning, and
  phase transitions in {{2D}} materials},\ }\href
  {https://doi.org/10.1038/s41699-021-00211-6} {\bibfield  {journal} {\bibinfo
  {journal} {npj 2D Mater. Appl.}\ }\textbf {\bibinfo {volume} {5}},\ \bibinfo
  {pages} {1} (\bibinfo {year} {2021})}\BibitemShut {NoStop}%
\bibitem [{\citenamefont {Wang}\ \emph {et~al.}(2024)\citenamefont {Wang},
  \citenamefont {Zheng}, \citenamefont {Chen}, \citenamefont {Ma},
  \citenamefont {Hong}, \citenamefont {Rodriguez}, \citenamefont {Woehl},
  \citenamefont {Shi}, \citenamefont {Parker},\ and\ \citenamefont
  {Ren}}]{wang2024}%
  \BibitemOpen
  \bibfield  {author} {\bibinfo {author} {\bibfnamefont {Z.}~\bibnamefont
  {Wang}}, \bibinfo {author} {\bibfnamefont {H.}~\bibnamefont {Zheng}},
  \bibinfo {author} {\bibfnamefont {A.}~\bibnamefont {Chen}}, \bibinfo {author}
  {\bibfnamefont {L.}~\bibnamefont {Ma}}, \bibinfo {author} {\bibfnamefont
  {S.~J.}\ \bibnamefont {Hong}}, \bibinfo {author} {\bibfnamefont {E.~E.}\
  \bibnamefont {Rodriguez}}, \bibinfo {author} {\bibfnamefont {T.~J.}\
  \bibnamefont {Woehl}}, \bibinfo {author} {\bibfnamefont {S.-F.}\ \bibnamefont
  {Shi}}, \bibinfo {author} {\bibfnamefont {T.}~\bibnamefont {Parker}},\ and\
  \bibinfo {author} {\bibfnamefont {S.}~\bibnamefont {Ren}},\ }\bibfield
  {title} {\bibinfo {title} {Room-{{Temperature CrI}}{\textsubscript{3}}
  {{Magnets}} through {{Lithiation}}},\ }\href
  {https://doi.org/10.1021/acsnano.4c02613} {\bibfield  {journal} {\bibinfo
  {journal} {ACS Nano}\ }\textbf {\bibinfo {volume} {18}},\ \bibinfo {pages}
  {23058} (\bibinfo {year} {2024})}\BibitemShut {NoStop}%
\bibitem [{\citenamefont {Zhang}\ \emph {et~al.}(2025)\citenamefont {Zhang},
  \citenamefont {Lv}, \citenamefont {Xu}, \citenamefont {Qi}, \citenamefont
  {Wang}, \citenamefont {Li}, \citenamefont {Su}, \citenamefont {Jiang},\ and\
  \citenamefont {Guan}}]{zhang2025}%
  \BibitemOpen
  \bibfield  {author} {\bibinfo {author} {\bibfnamefont {F.}~\bibnamefont
  {Zhang}}, \bibinfo {author} {\bibfnamefont {L.}~\bibnamefont {Lv}}, \bibinfo
  {author} {\bibfnamefont {Z.}~\bibnamefont {Xu}}, \bibinfo {author}
  {\bibfnamefont {D.}~\bibnamefont {Qi}}, \bibinfo {author} {\bibfnamefont
  {W.}~\bibnamefont {Wang}}, \bibinfo {author} {\bibfnamefont {X.}~\bibnamefont
  {Li}}, \bibinfo {author} {\bibfnamefont {Y.}~\bibnamefont {Su}}, \bibinfo
  {author} {\bibfnamefont {Y.}~\bibnamefont {Jiang}},\ and\ \bibinfo {author}
  {\bibfnamefont {Z.}~\bibnamefont {Guan}},\ }\bibfield  {title} {\bibinfo
  {title} {Prediction of the {{TiS{\textsubscript{2}} Bilayer}} with
  {{Self-Intercalation}}: {{Robust Ferromagnetic Semiconductor}} with a {{High
  Curie Temperature}}},\ }\href {https://doi.org/10.1021/acs.jpcc.4c06216}
  {\bibfield  {journal} {\bibinfo  {journal} {J. Phys. Chem. C}\ }\textbf
  {\bibinfo {volume} {129}},\ \bibinfo {pages} {5577} (\bibinfo {year}
  {2025})}\BibitemShut {NoStop}%
\bibitem [{\citenamefont {Liu}\ \emph {et~al.}(2024)\citenamefont {Liu},
  \citenamefont {Li}, \citenamefont {Chen}, \citenamefont {Hu}, \citenamefont
  {Duan}, \citenamefont {Wang}, \citenamefont {Feng}, \citenamefont {Liu},
  \citenamefont {Zhang}, \citenamefont {Cao}, \citenamefont {Niu},
  \citenamefont {Li}, \citenamefont {Li},\ and\ \citenamefont {Yan}}]{liu2024}%
  \BibitemOpen
  \bibfield  {author} {\bibinfo {author} {\bibfnamefont {C.}~\bibnamefont
  {Liu}}, \bibinfo {author} {\bibfnamefont {Z.}~\bibnamefont {Li}}, \bibinfo
  {author} {\bibfnamefont {Z.}~\bibnamefont {Chen}}, \bibinfo {author}
  {\bibfnamefont {J.}~\bibnamefont {Hu}}, \bibinfo {author} {\bibfnamefont
  {H.}~\bibnamefont {Duan}}, \bibinfo {author} {\bibfnamefont {C.}~\bibnamefont
  {Wang}}, \bibinfo {author} {\bibfnamefont {S.}~\bibnamefont {Feng}}, \bibinfo
  {author} {\bibfnamefont {R.}~\bibnamefont {Liu}}, \bibinfo {author}
  {\bibfnamefont {G.}~\bibnamefont {Zhang}}, \bibinfo {author} {\bibfnamefont
  {J.}~\bibnamefont {Cao}}, \bibinfo {author} {\bibfnamefont {Y.}~\bibnamefont
  {Niu}}, \bibinfo {author} {\bibfnamefont {Q.}~\bibnamefont {Li}}, \bibinfo
  {author} {\bibfnamefont {P.}~\bibnamefont {Li}},\ and\ \bibinfo {author}
  {\bibfnamefont {W.}~\bibnamefont {Yan}},\ }\bibfield  {title} {\bibinfo
  {title} {Realizing {{Room}}-{{Temperature Ferromagnetism}} in
  {{Molecular}}-{{Intercalated Antiferromagnet VOCl}}},\ }\href
  {https://doi.org/10.1002/adma.202405284} {\bibfield  {journal} {\bibinfo
  {journal} {Adv. Mater.}\ }\textbf {\bibinfo {volume} {36}},\ \bibinfo {pages}
  {2405284} (\bibinfo {year} {2024})}\BibitemShut {NoStop}%
\bibitem [{\citenamefont {Kappera}\ \emph {et~al.}(2014)\citenamefont
  {Kappera}, \citenamefont {Voiry}, \citenamefont {Yalcin}, \citenamefont
  {Branch}, \citenamefont {Gupta}, \citenamefont {Mohite},\ and\ \citenamefont
  {Chhowalla}}]{kappera2014}%
  \BibitemOpen
  \bibfield  {author} {\bibinfo {author} {\bibfnamefont {R.}~\bibnamefont
  {Kappera}}, \bibinfo {author} {\bibfnamefont {D.}~\bibnamefont {Voiry}},
  \bibinfo {author} {\bibfnamefont {S.~E.}\ \bibnamefont {Yalcin}}, \bibinfo
  {author} {\bibfnamefont {B.}~\bibnamefont {Branch}}, \bibinfo {author}
  {\bibfnamefont {G.}~\bibnamefont {Gupta}}, \bibinfo {author} {\bibfnamefont
  {A.~D.}\ \bibnamefont {Mohite}},\ and\ \bibinfo {author} {\bibfnamefont
  {M.}~\bibnamefont {Chhowalla}},\ }\bibfield  {title} {\bibinfo {title}
  {Phase-engineered low-resistance contacts for ultrathin
  {{MoS{\textsubscript{2}}}} transistors},\ }\href
  {https://doi.org/10.1038/nmat4080} {\bibfield  {journal} {\bibinfo  {journal}
  {Nat. Mater.}\ }\textbf {\bibinfo {volume} {13}},\ \bibinfo {pages} {1128}
  (\bibinfo {year} {2014})}\BibitemShut {NoStop}%
\bibitem [{\citenamefont {Yang}\ \emph {et~al.}(2022)\citenamefont {Yang},
  \citenamefont {Mei}, \citenamefont {Zhang}, \citenamefont {Fan},
  \citenamefont {Shin}, \citenamefont {Voiry},\ and\ \citenamefont
  {Zeng}}]{yangReview2022}%
  \BibitemOpen
  \bibfield  {author} {\bibinfo {author} {\bibfnamefont {R.}~\bibnamefont
  {Yang}}, \bibinfo {author} {\bibfnamefont {L.}~\bibnamefont {Mei}}, \bibinfo
  {author} {\bibfnamefont {Q.}~\bibnamefont {Zhang}}, \bibinfo {author}
  {\bibfnamefont {Y.}~\bibnamefont {Fan}}, \bibinfo {author} {\bibfnamefont
  {H.~S.}\ \bibnamefont {Shin}}, \bibinfo {author} {\bibfnamefont
  {D.}~\bibnamefont {Voiry}},\ and\ \bibinfo {author} {\bibfnamefont
  {Z.}~\bibnamefont {Zeng}},\ }\bibfield  {title} {\bibinfo {title} {High-yield
  production of mono- or few-layer transition metal dichalcogenide nanosheets
  by an electrochemical lithium ion intercalation-based exfoliation method},\
  }\href {https://doi.org/10.1038/s41596-021-00643-w} {\bibfield  {journal}
  {\bibinfo  {journal} {Nat. Protoc.}\ }\textbf {\bibinfo {volume} {17}},\
  \bibinfo {pages} {358} (\bibinfo {year} {2022})}\BibitemShut {NoStop}%
\bibitem [{\citenamefont {Yang}\ \emph {et~al.}(2023)\citenamefont {Yang},
  \citenamefont {Fan}, \citenamefont {Mei}, \citenamefont {Shin}, \citenamefont
  {Voiry}, \citenamefont {Lu}, \citenamefont {Li},\ and\ \citenamefont
  {Zeng}}]{yangReview2023}%
  \BibitemOpen
  \bibfield  {author} {\bibinfo {author} {\bibfnamefont {R.}~\bibnamefont
  {Yang}}, \bibinfo {author} {\bibfnamefont {Y.}~\bibnamefont {Fan}}, \bibinfo
  {author} {\bibfnamefont {L.}~\bibnamefont {Mei}}, \bibinfo {author}
  {\bibfnamefont {H.~S.}\ \bibnamefont {Shin}}, \bibinfo {author}
  {\bibfnamefont {D.}~\bibnamefont {Voiry}}, \bibinfo {author} {\bibfnamefont
  {Q.}~\bibnamefont {Lu}}, \bibinfo {author} {\bibfnamefont {J.}~\bibnamefont
  {Li}},\ and\ \bibinfo {author} {\bibfnamefont {Z.}~\bibnamefont {Zeng}},\
  }\bibfield  {title} {\bibinfo {title} {Synthesis of atomically thin sheets by
  the intercalation-based exfoliation of layered materials},\ }\href
  {https://doi.org/10.1038/s44160-022-00232-z} {\bibfield  {journal} {\bibinfo
  {journal} {Nat. Synth.}\ }\textbf {\bibinfo {volume} {2}},\ \bibinfo {pages}
  {101} (\bibinfo {year} {2023})}\BibitemShut {NoStop}%
\bibitem [{\citenamefont {Yang}\ \emph {et~al.}(2024)\citenamefont {Yang},
  \citenamefont {Mei}, \citenamefont {Lin}, \citenamefont {Fan}, \citenamefont
  {Lim}, \citenamefont {Guo}, \citenamefont {Liu}, \citenamefont {Shin},
  \citenamefont {Voiry}, \citenamefont {Lu}, \citenamefont {Li},\ and\
  \citenamefont {Zeng}}]{yangReview2024a}%
  \BibitemOpen
  \bibfield  {author} {\bibinfo {author} {\bibfnamefont {R.}~\bibnamefont
  {Yang}}, \bibinfo {author} {\bibfnamefont {L.}~\bibnamefont {Mei}}, \bibinfo
  {author} {\bibfnamefont {Z.}~\bibnamefont {Lin}}, \bibinfo {author}
  {\bibfnamefont {Y.}~\bibnamefont {Fan}}, \bibinfo {author} {\bibfnamefont
  {J.}~\bibnamefont {Lim}}, \bibinfo {author} {\bibfnamefont {J.}~\bibnamefont
  {Guo}}, \bibinfo {author} {\bibfnamefont {Y.}~\bibnamefont {Liu}}, \bibinfo
  {author} {\bibfnamefont {H.~S.}\ \bibnamefont {Shin}}, \bibinfo {author}
  {\bibfnamefont {D.}~\bibnamefont {Voiry}}, \bibinfo {author} {\bibfnamefont
  {Q.}~\bibnamefont {Lu}}, \bibinfo {author} {\bibfnamefont {J.}~\bibnamefont
  {Li}},\ and\ \bibinfo {author} {\bibfnamefont {Z.}~\bibnamefont {Zeng}},\
  }\bibfield  {title} {\bibinfo {title} {Intercalation in {{2D}} materials and
  in situ studies},\ }\href {https://doi.org/10.1038/s41570-024-00605-2}
  {\bibfield  {journal} {\bibinfo  {journal} {Nat. Rev. Chem.}\ }\textbf
  {\bibinfo {volume} {8}},\ \bibinfo {pages} {410} (\bibinfo {year}
  {2024})}\BibitemShut {NoStop}%
\bibitem [{\citenamefont {Zhu}\ \emph {et~al.}(2019)\citenamefont {Zhu},
  \citenamefont {Li}, \citenamefont {Liang},\ and\ \citenamefont
  {Lu}}]{zhu2019}%
  \BibitemOpen
  \bibfield  {author} {\bibinfo {author} {\bibfnamefont {X.}~\bibnamefont
  {Zhu}}, \bibinfo {author} {\bibfnamefont {D.}~\bibnamefont {Li}}, \bibinfo
  {author} {\bibfnamefont {X.}~\bibnamefont {Liang}},\ and\ \bibinfo {author}
  {\bibfnamefont {W.~D.}\ \bibnamefont {Lu}},\ }\bibfield  {title} {\bibinfo
  {title} {Ionic modulation and ionic coupling effects in
  {{MoS{\textsubscript{2}}}} devices for neuromorphic computing},\ }\href
  {https://doi.org/10.1038/s41563-018-0248-5} {\bibfield  {journal} {\bibinfo
  {journal} {Nat. Mater.}\ }\textbf {\bibinfo {volume} {18}},\ \bibinfo {pages}
  {141} (\bibinfo {year} {2019})}\BibitemShut {NoStop}%
\bibitem [{\citenamefont {Kanetani}\ \emph {et~al.}(2012)\citenamefont
  {Kanetani}, \citenamefont {Sugawara}, \citenamefont {Sato}, \citenamefont
  {Shimizu}, \citenamefont {Iwaya}, \citenamefont {Hitosugi},\ and\
  \citenamefont {Takahashi}}]{kanetani2012}%
  \BibitemOpen
  \bibfield  {author} {\bibinfo {author} {\bibfnamefont {K.}~\bibnamefont
  {Kanetani}}, \bibinfo {author} {\bibfnamefont {K.}~\bibnamefont {Sugawara}},
  \bibinfo {author} {\bibfnamefont {T.}~\bibnamefont {Sato}}, \bibinfo {author}
  {\bibfnamefont {R.}~\bibnamefont {Shimizu}}, \bibinfo {author} {\bibfnamefont
  {K.}~\bibnamefont {Iwaya}}, \bibinfo {author} {\bibfnamefont
  {T.}~\bibnamefont {Hitosugi}},\ and\ \bibinfo {author} {\bibfnamefont
  {T.}~\bibnamefont {Takahashi}},\ }\bibfield  {title} {\bibinfo {title} {Ca
  intercalated bilayer graphene as a thinnest limit of superconducting
  {{C{\textsubscript{6}}Ca}}},\ }\href
  {https://doi.org/10.1073/pnas.1208889109} {\bibfield  {journal} {\bibinfo
  {journal} {Proc. Natl. Acad. Sci.}\ }\textbf {\bibinfo {volume} {109}},\
  \bibinfo {pages} {19610} (\bibinfo {year} {2012})}\BibitemShut {NoStop}%
\bibitem [{\citenamefont {Xiong}\ \emph {et~al.}(2015)\citenamefont {Xiong},
  \citenamefont {Wang}, \citenamefont {Liu}, \citenamefont {Sun}, \citenamefont
  {Brongersma}, \citenamefont {Pop},\ and\ \citenamefont {Cui}}]{xiongLi2015}%
  \BibitemOpen
  \bibfield  {author} {\bibinfo {author} {\bibfnamefont {F.}~\bibnamefont
  {Xiong}}, \bibinfo {author} {\bibfnamefont {H.}~\bibnamefont {Wang}},
  \bibinfo {author} {\bibfnamefont {X.}~\bibnamefont {Liu}}, \bibinfo {author}
  {\bibfnamefont {J.}~\bibnamefont {Sun}}, \bibinfo {author} {\bibfnamefont
  {M.}~\bibnamefont {Brongersma}}, \bibinfo {author} {\bibfnamefont
  {E.}~\bibnamefont {Pop}},\ and\ \bibinfo {author} {\bibfnamefont
  {Y.}~\bibnamefont {Cui}},\ }\bibfield  {title} {\bibinfo {title} {Li
  {{Intercalation}} in {{MoS{\textsubscript{2}}}}: {{In Situ Observation}} of
  {{Its Dynamics}} and {{Tuning Optical}} and {{Electrical Properties}}},\
  }\href {https://doi.org/10.1021/acs.nanolett.5b02619} {\bibfield  {journal}
  {\bibinfo  {journal} {Nano Lett.}\ }\textbf {\bibinfo {volume} {15}},\
  \bibinfo {pages} {6777} (\bibinfo {year} {2015})}\BibitemShut {NoStop}%
\bibitem [{\citenamefont {Yao}\ \emph {et~al.}(2014)\citenamefont {Yao},
  \citenamefont {Koski}, \citenamefont {Luo}, \citenamefont {Cha},
  \citenamefont {Hu}, \citenamefont {Kong}, \citenamefont {Narasimhan},
  \citenamefont {Huo},\ and\ \citenamefont {Cui}}]{yaoCu2014}%
  \BibitemOpen
  \bibfield  {author} {\bibinfo {author} {\bibfnamefont {J.}~\bibnamefont
  {Yao}}, \bibinfo {author} {\bibfnamefont {K.~J.}\ \bibnamefont {Koski}},
  \bibinfo {author} {\bibfnamefont {W.}~\bibnamefont {Luo}}, \bibinfo {author}
  {\bibfnamefont {J.~J.}\ \bibnamefont {Cha}}, \bibinfo {author} {\bibfnamefont
  {L.}~\bibnamefont {Hu}}, \bibinfo {author} {\bibfnamefont {D.}~\bibnamefont
  {Kong}}, \bibinfo {author} {\bibfnamefont {V.~K.}\ \bibnamefont
  {Narasimhan}}, \bibinfo {author} {\bibfnamefont {K.}~\bibnamefont {Huo}},\
  and\ \bibinfo {author} {\bibfnamefont {Y.}~\bibnamefont {Cui}},\ }\bibfield
  {title} {\bibinfo {title} {Optical transmission enhacement through chemically
  tuned two-dimensional bismuth chalcogenide nanoplates},\ }\href
  {https://doi.org/10.1038/ncomms6670} {\bibfield  {journal} {\bibinfo
  {journal} {Nat. Commun.}\ }\textbf {\bibinfo {volume} {5}},\ \bibinfo {pages}
  {5670} (\bibinfo {year} {2014})}\BibitemShut {NoStop}%
\bibitem [{\citenamefont {Zhang}\ \emph {et~al.}(2022)\citenamefont {Zhang},
  \citenamefont {Wei}, \citenamefont {Kang}, \citenamefont {Pu}, \citenamefont
  {Li}, \citenamefont {Ma}, \citenamefont {Xu},\ and\ \citenamefont
  {Luo}}]{zhang2022}%
  \BibitemOpen
  \bibfield  {author} {\bibinfo {author} {\bibfnamefont {R.}~\bibnamefont
  {Zhang}}, \bibinfo {author} {\bibfnamefont {Y.}~\bibnamefont {Wei}}, \bibinfo
  {author} {\bibfnamefont {Y.}~\bibnamefont {Kang}}, \bibinfo {author}
  {\bibfnamefont {M.}~\bibnamefont {Pu}}, \bibinfo {author} {\bibfnamefont
  {X.}~\bibnamefont {Li}}, \bibinfo {author} {\bibfnamefont {X.}~\bibnamefont
  {Ma}}, \bibinfo {author} {\bibfnamefont {M.}~\bibnamefont {Xu}},\ and\
  \bibinfo {author} {\bibfnamefont {X.}~\bibnamefont {Luo}},\ }\bibfield
  {title} {\bibinfo {title} {Breaking the {{Cut-Off Wavelength Limit}} of
  {{GaTe}} through {{Self-Driven Oxygen Intercalation}} in {{Air}}},\ }\href
  {https://doi.org/10.1002/advs.202103429} {\bibfield  {journal} {\bibinfo
  {journal} {Adv. Sci.}\ }\textbf {\bibinfo {volume} {9}},\ \bibinfo {pages}
  {2103429} (\bibinfo {year} {2022})}\BibitemShut {NoStop}%
\bibitem [{\citenamefont {Feuer}\ \emph {et~al.}(2025)\citenamefont {Feuer},
  \citenamefont {Thinel}, \citenamefont {Huang}, \citenamefont {Cui},
  \citenamefont {Shao}, \citenamefont {Kundu}, \citenamefont {Chica},
  \citenamefont {Han}, \citenamefont {Pokratath}, \citenamefont {Telford},
  \citenamefont {Cox}, \citenamefont {York}, \citenamefont {Okuno},
  \citenamefont {Huang}, \citenamefont {Bukula}, \citenamefont {Nashabeh},
  \citenamefont {Qiu}, \citenamefont {Nuckolls}, \citenamefont {Dean},
  \citenamefont {Billinge}, \citenamefont {Zhu}, \citenamefont {Zhu},
  \citenamefont {Basov}, \citenamefont {Millis}, \citenamefont {Reichman},
  \citenamefont {Pasupathy}, \citenamefont {Roy},\ and\ \citenamefont
  {Ziebel}}]{feuer2025}%
  \BibitemOpen
  \bibfield  {author} {\bibinfo {author} {\bibfnamefont {M.~L.}\ \bibnamefont
  {Feuer}}, \bibinfo {author} {\bibfnamefont {M.}~\bibnamefont {Thinel}},
  \bibinfo {author} {\bibfnamefont {X.}~\bibnamefont {Huang}}, \bibinfo
  {author} {\bibfnamefont {Z.-H.}\ \bibnamefont {Cui}}, \bibinfo {author}
  {\bibfnamefont {Y.}~\bibnamefont {Shao}}, \bibinfo {author} {\bibfnamefont
  {A.~K.}\ \bibnamefont {Kundu}}, \bibinfo {author} {\bibfnamefont {D.~G.}\
  \bibnamefont {Chica}}, \bibinfo {author} {\bibfnamefont {M.-G.}\ \bibnamefont
  {Han}}, \bibinfo {author} {\bibfnamefont {R.}~\bibnamefont {Pokratath}},
  \bibinfo {author} {\bibfnamefont {E.~J.}\ \bibnamefont {Telford}}, \bibinfo
  {author} {\bibfnamefont {J.}~\bibnamefont {Cox}}, \bibinfo {author}
  {\bibfnamefont {E.}~\bibnamefont {York}}, \bibinfo {author} {\bibfnamefont
  {S.}~\bibnamefont {Okuno}}, \bibinfo {author} {\bibfnamefont {C.-Y.}\
  \bibnamefont {Huang}}, \bibinfo {author} {\bibfnamefont {O.}~\bibnamefont
  {Bukula}}, \bibinfo {author} {\bibfnamefont {L.~M.}\ \bibnamefont
  {Nashabeh}}, \bibinfo {author} {\bibfnamefont {S.}~\bibnamefont {Qiu}},
  \bibinfo {author} {\bibfnamefont {C.~P.}\ \bibnamefont {Nuckolls}}, \bibinfo
  {author} {\bibfnamefont {C.~R.}\ \bibnamefont {Dean}}, \bibinfo {author}
  {\bibfnamefont {S.~J.~L.}\ \bibnamefont {Billinge}}, \bibinfo {author}
  {\bibfnamefont {X.}~\bibnamefont {Zhu}}, \bibinfo {author} {\bibfnamefont
  {Y.}~\bibnamefont {Zhu}}, \bibinfo {author} {\bibfnamefont {D.~N.}\
  \bibnamefont {Basov}}, \bibinfo {author} {\bibfnamefont {A.~J.}\ \bibnamefont
  {Millis}}, \bibinfo {author} {\bibfnamefont {D.~R.}\ \bibnamefont
  {Reichman}}, \bibinfo {author} {\bibfnamefont {A.~N.}\ \bibnamefont
  {Pasupathy}}, \bibinfo {author} {\bibfnamefont {X.}~\bibnamefont {Roy}},\
  and\ \bibinfo {author} {\bibfnamefont {M.~E.}\ \bibnamefont {Ziebel}},\
  }\bibfield  {title} {\bibinfo {title} {Charge {{Density Wave}} and
  {{Ferromagnetism}} in {{Intercalated CrSBr}}},\ }\href
  {https://doi.org/10.1002/adma.202418066} {\bibfield  {journal} {\bibinfo
  {journal} {Adv. Mater.}\ }\textbf {\bibinfo {volume} {37}},\ \bibinfo {pages}
  {2418066} (\bibinfo {year} {2025})}\BibitemShut {NoStop}%
\bibitem [{\citenamefont {Huan}\ \emph {et~al.}(2023)\citenamefont {Huan},
  \citenamefont {Luo}, \citenamefont {Han}, \citenamefont {Ge}, \citenamefont
  {Cui}, \citenamefont {Zhu}, \citenamefont {Hu}, \citenamefont {Zheng},
  \citenamefont {Zhao}, \citenamefont {Wang}, \citenamefont {Wang},\ and\
  \citenamefont {Zhang}}]{huanFe2023}%
  \BibitemOpen
  \bibfield  {author} {\bibinfo {author} {\bibfnamefont {Y.}~\bibnamefont
  {Huan}}, \bibinfo {author} {\bibfnamefont {T.}~\bibnamefont {Luo}}, \bibinfo
  {author} {\bibfnamefont {X.}~\bibnamefont {Han}}, \bibinfo {author}
  {\bibfnamefont {J.}~\bibnamefont {Ge}}, \bibinfo {author} {\bibfnamefont
  {F.}~\bibnamefont {Cui}}, \bibinfo {author} {\bibfnamefont {L.}~\bibnamefont
  {Zhu}}, \bibinfo {author} {\bibfnamefont {J.}~\bibnamefont {Hu}}, \bibinfo
  {author} {\bibfnamefont {F.}~\bibnamefont {Zheng}}, \bibinfo {author}
  {\bibfnamefont {X.}~\bibnamefont {Zhao}}, \bibinfo {author} {\bibfnamefont
  {L.}~\bibnamefont {Wang}}, \bibinfo {author} {\bibfnamefont {J.}~\bibnamefont
  {Wang}},\ and\ \bibinfo {author} {\bibfnamefont {Y.}~\bibnamefont {Zhang}},\
  }\bibfield  {title} {\bibinfo {title} {Composition-{{Controllable Syntheses}}
  and {{Property Modulations}} from {{2D Ferromagnetic
  Fe{\textsubscript{5}}Se{\textsubscript{8}}}} to {{Metallic
  Fe{\textsubscript{3}}Se{\textsubscript{4}} Nanosheets}}},\ }\href
  {https://doi.org/10.1002/adma.202207276} {\bibfield  {journal} {\bibinfo
  {journal} {Adv. Mater.}\ }\textbf {\bibinfo {volume} {35}},\ \bibinfo {pages}
  {2207276} (\bibinfo {year} {2023})}\BibitemShut {NoStop}%
\bibitem [{\citenamefont {Iturriaga}\ \emph {et~al.}(2023)\citenamefont
  {Iturriaga}, \citenamefont {Martinez}, \citenamefont {Mai}, \citenamefont
  {Biacchi}, \citenamefont {Augustin}, \citenamefont {Hight~Walker},
  \citenamefont {Sanad}, \citenamefont {Sreenivasan}, \citenamefont {Liu},
  \citenamefont {Santos}, \citenamefont {Petrovic},\ and\ \citenamefont
  {Singamaneni}}]{iturriaga2023}%
  \BibitemOpen
  \bibfield  {author} {\bibinfo {author} {\bibfnamefont {H.}~\bibnamefont
  {Iturriaga}}, \bibinfo {author} {\bibfnamefont {L.~M.}\ \bibnamefont
  {Martinez}}, \bibinfo {author} {\bibfnamefont {T.~T.}\ \bibnamefont {Mai}},
  \bibinfo {author} {\bibfnamefont {A.~J.}\ \bibnamefont {Biacchi}}, \bibinfo
  {author} {\bibfnamefont {M.}~\bibnamefont {Augustin}}, \bibinfo {author}
  {\bibfnamefont {A.~R.}\ \bibnamefont {Hight~Walker}}, \bibinfo {author}
  {\bibfnamefont {M.~F.}\ \bibnamefont {Sanad}}, \bibinfo {author}
  {\bibfnamefont {S.~T.}\ \bibnamefont {Sreenivasan}}, \bibinfo {author}
  {\bibfnamefont {Y.}~\bibnamefont {Liu}}, \bibinfo {author} {\bibfnamefont
  {E.~J.~G.}\ \bibnamefont {Santos}}, \bibinfo {author} {\bibfnamefont
  {C.}~\bibnamefont {Petrovic}},\ and\ \bibinfo {author} {\bibfnamefont
  {S.~R.}\ \bibnamefont {Singamaneni}},\ }\bibfield  {title} {\bibinfo {title}
  {Magnetic properties of intercalated quasi-{{2D
  Fe{\textsubscript{3-x}}GeTe{\textsubscript{2}}}} van der {{Waals}} magnet},\
  }\href {https://doi.org/10.1038/s41699-023-00417-w} {\bibfield  {journal}
  {\bibinfo  {journal} {npj 2D Mater. Appl.}\ }\textbf {\bibinfo {volume}
  {7}},\ \bibinfo {pages} {1} (\bibinfo {year} {2023})}\BibitemShut {NoStop}%
\bibitem [{\citenamefont {Mishra}\ \emph {et~al.}(2024)\citenamefont {Mishra},
  \citenamefont {Park}, \citenamefont {Javaid}, \citenamefont {Shin},\ and\
  \citenamefont {Lee}}]{mishra2024}%
  \BibitemOpen
  \bibfield  {author} {\bibinfo {author} {\bibfnamefont {S.}~\bibnamefont
  {Mishra}}, \bibinfo {author} {\bibfnamefont {I.~K.}\ \bibnamefont {Park}},
  \bibinfo {author} {\bibfnamefont {S.}~\bibnamefont {Javaid}}, \bibinfo
  {author} {\bibfnamefont {S.~H.}\ \bibnamefont {Shin}},\ and\ \bibinfo
  {author} {\bibfnamefont {G.}~\bibnamefont {Lee}},\ }\bibfield  {title}
  {\bibinfo {title} {Enhancement of interlayer exchange coupling via
  intercalation in {{2D}} magnetic bilayers: Towards high {{Curie}}
  temperature},\ }\href {https://doi.org/10.1039/D4MH00135D} {\bibfield
  {journal} {\bibinfo  {journal} {Mater. Horiz.}\ }\textbf {\bibinfo {volume}
  {11}},\ \bibinfo {pages} {4482} (\bibinfo {year} {2024})}\BibitemShut
  {NoStop}%
\bibitem [{\citenamefont {Mi}\ \emph {et~al.}(2022)\citenamefont {Mi},
  \citenamefont {Zheng}, \citenamefont {Wang}, \citenamefont {Zhou},
  \citenamefont {Yu}, \citenamefont {Xiao}, \citenamefont {Song}, \citenamefont
  {Shen}, \citenamefont {Li}, \citenamefont {Bai}, \citenamefont {Chen},
  \citenamefont {Wang}, \citenamefont {Liu},\ and\ \citenamefont
  {Wang}}]{mi2022}%
  \BibitemOpen
  \bibfield  {author} {\bibinfo {author} {\bibfnamefont {M.}~\bibnamefont
  {Mi}}, \bibinfo {author} {\bibfnamefont {X.}~\bibnamefont {Zheng}}, \bibinfo
  {author} {\bibfnamefont {S.}~\bibnamefont {Wang}}, \bibinfo {author}
  {\bibfnamefont {Y.}~\bibnamefont {Zhou}}, \bibinfo {author} {\bibfnamefont
  {L.}~\bibnamefont {Yu}}, \bibinfo {author} {\bibfnamefont {H.}~\bibnamefont
  {Xiao}}, \bibinfo {author} {\bibfnamefont {H.}~\bibnamefont {Song}}, \bibinfo
  {author} {\bibfnamefont {B.}~\bibnamefont {Shen}}, \bibinfo {author}
  {\bibfnamefont {F.}~\bibnamefont {Li}}, \bibinfo {author} {\bibfnamefont
  {L.}~\bibnamefont {Bai}}, \bibinfo {author} {\bibfnamefont {Y.}~\bibnamefont
  {Chen}}, \bibinfo {author} {\bibfnamefont {S.}~\bibnamefont {Wang}}, \bibinfo
  {author} {\bibfnamefont {X.}~\bibnamefont {Liu}},\ and\ \bibinfo {author}
  {\bibfnamefont {Y.}~\bibnamefont {Wang}},\ }\bibfield  {title} {\bibinfo
  {title} {Variation between {{Antiferromagnetism}} and {{Ferrimagnetism}} in
  {{NiPS}}{\textsubscript{3}} by {{Electron Doping}}},\ }\href
  {https://doi.org/10.1002/adfm.202112750} {\bibfield  {journal} {\bibinfo
  {journal} {Adv. Funct. Mater.}\ }\textbf {\bibinfo {volume} {32}},\ \bibinfo
  {pages} {2112750} (\bibinfo {year} {2022})}\BibitemShut {NoStop}%
\bibitem [{\citenamefont {Weber}\ \emph {et~al.}(2019)\citenamefont {Weber},
  \citenamefont {Trout}, \citenamefont {McComb},\ and\ \citenamefont
  {Goldberger}}]{weber2019}%
  \BibitemOpen
  \bibfield  {author} {\bibinfo {author} {\bibfnamefont {D.}~\bibnamefont
  {Weber}}, \bibinfo {author} {\bibfnamefont {A.~H.}\ \bibnamefont {Trout}},
  \bibinfo {author} {\bibfnamefont {D.~W.}\ \bibnamefont {McComb}},\ and\
  \bibinfo {author} {\bibfnamefont {J.~E.}\ \bibnamefont {Goldberger}},\
  }\bibfield  {title} {\bibinfo {title} {Decomposition-{{Induced
  Room-Temperature Magnetism}} of the {{Na-Intercalated Layered Ferromagnet
  Fe}}{\textsubscript{3-x}}{{GeTe}}{\textsubscript{2}}},\ }\href
  {https://doi.org/10.1021/acs.nanolett.9b01287} {\bibfield  {journal}
  {\bibinfo  {journal} {Nano Lett.}\ }\textbf {\bibinfo {volume} {19}},\
  \bibinfo {pages} {5031} (\bibinfo {year} {2019})}\BibitemShut {NoStop}%
\bibitem [{\citenamefont {Wu}\ \emph {et~al.}(2023)\citenamefont {Wu},
  \citenamefont {Hu}, \citenamefont {Wang}, \citenamefont {Zhou}, \citenamefont
  {Hou}, \citenamefont {Xia}, \citenamefont {Zhang}, \citenamefont {Wang},
  \citenamefont {Ding}, \citenamefont {He}, \citenamefont {Dong}, \citenamefont
  {Bao}, \citenamefont {Wen}, \citenamefont {Guo}, \citenamefont {Watanabe},
  \citenamefont {Taniguchi}, \citenamefont {Ji}, \citenamefont {Wang},\ and\
  \citenamefont {Li}}]{wuFe2023}%
  \BibitemOpen
  \bibfield  {author} {\bibinfo {author} {\bibfnamefont {Y.}~\bibnamefont
  {Wu}}, \bibinfo {author} {\bibfnamefont {Y.}~\bibnamefont {Hu}}, \bibinfo
  {author} {\bibfnamefont {C.}~\bibnamefont {Wang}}, \bibinfo {author}
  {\bibfnamefont {X.}~\bibnamefont {Zhou}}, \bibinfo {author} {\bibfnamefont
  {X.}~\bibnamefont {Hou}}, \bibinfo {author} {\bibfnamefont {W.}~\bibnamefont
  {Xia}}, \bibinfo {author} {\bibfnamefont {Y.}~\bibnamefont {Zhang}}, \bibinfo
  {author} {\bibfnamefont {J.}~\bibnamefont {Wang}}, \bibinfo {author}
  {\bibfnamefont {Y.}~\bibnamefont {Ding}}, \bibinfo {author} {\bibfnamefont
  {J.}~\bibnamefont {He}}, \bibinfo {author} {\bibfnamefont {P.}~\bibnamefont
  {Dong}}, \bibinfo {author} {\bibfnamefont {S.}~\bibnamefont {Bao}}, \bibinfo
  {author} {\bibfnamefont {J.}~\bibnamefont {Wen}}, \bibinfo {author}
  {\bibfnamefont {Y.}~\bibnamefont {Guo}}, \bibinfo {author} {\bibfnamefont
  {K.}~\bibnamefont {Watanabe}}, \bibinfo {author} {\bibfnamefont
  {T.}~\bibnamefont {Taniguchi}}, \bibinfo {author} {\bibfnamefont
  {W.}~\bibnamefont {Ji}}, \bibinfo {author} {\bibfnamefont {Z.-J.}\
  \bibnamefont {Wang}},\ and\ \bibinfo {author} {\bibfnamefont
  {J.}~\bibnamefont {Li}},\ }\bibfield  {title} {\bibinfo {title}
  {Fe-{{Intercalation Dominated Ferromagnetism}} of van der {{Waals
  Fe{\textsubscript{3}}GeTe{\textsubscript{2}}}}},\ }\href
  {https://doi.org/10.1002/adma.202302568} {\bibfield  {journal} {\bibinfo
  {journal} {Adv. Mater.}\ }\textbf {\bibinfo {volume} {35}},\ \bibinfo {pages}
  {2302568} (\bibinfo {year} {2023})}\BibitemShut {NoStop}%
\bibitem [{\citenamefont {Zhao}\ \emph {et~al.}(2024)\citenamefont {Zhao},
  \citenamefont {Gao}, \citenamefont {An}, \citenamefont {Xu}, \citenamefont
  {Tian},\ and\ \citenamefont {Xu}}]{zhaoCu2024}%
  \BibitemOpen
  \bibfield  {author} {\bibinfo {author} {\bibfnamefont {D.}~\bibnamefont
  {Zhao}}, \bibinfo {author} {\bibfnamefont {B.}~\bibnamefont {Gao}}, \bibinfo
  {author} {\bibfnamefont {G.}~\bibnamefont {An}}, \bibinfo {author}
  {\bibfnamefont {S.}~\bibnamefont {Xu}}, \bibinfo {author} {\bibfnamefont
  {Q.}~\bibnamefont {Tian}},\ and\ \bibinfo {author} {\bibfnamefont
  {Q.}~\bibnamefont {Xu}},\ }\bibfield  {title} {\bibinfo {title} {Copper
  {{Intercalation Induces Amorphization}} of {{2D
  Cu}}/{{WO{\textsubscript{3}}}} for {{Room-Temperature Ferromagnetism}}},\
  }\href {https://doi.org/10.1002/anie.202412811} {\bibfield  {journal}
  {\bibinfo  {journal} {Angew. Chem. Int. Ed.}\ }\textbf {\bibinfo {volume}
  {63}},\ \bibinfo {pages} {e202412811} (\bibinfo {year} {2024})}\BibitemShut
  {NoStop}%
\bibitem [{\citenamefont {Bensch}\ \emph {et~al.}(2009)\citenamefont {Bensch},
  \citenamefont {Bredow}, \citenamefont {Ebert}, \citenamefont {Heitjans},
  \citenamefont {Indris}, \citenamefont {Mankovsky},\ and\ \citenamefont
  {Wilkening}}]{bensch2009}%
  \BibitemOpen
  \bibfield  {author} {\bibinfo {author} {\bibfnamefont {W.}~\bibnamefont
  {Bensch}}, \bibinfo {author} {\bibfnamefont {T.}~\bibnamefont {Bredow}},
  \bibinfo {author} {\bibfnamefont {H.}~\bibnamefont {Ebert}}, \bibinfo
  {author} {\bibfnamefont {P.}~\bibnamefont {Heitjans}}, \bibinfo {author}
  {\bibfnamefont {S.}~\bibnamefont {Indris}}, \bibinfo {author} {\bibfnamefont
  {S.}~\bibnamefont {Mankovsky}},\ and\ \bibinfo {author} {\bibfnamefont
  {M.}~\bibnamefont {Wilkening}},\ }\bibfield  {title} {\bibinfo {title} {Li
  intercalation and anion/cation substitution of transition metal
  chalcogenides: {{Effects}} on crystal structure, microstructure, magnetic
  properties and {{Li}}+ ion mobility},\ }\href
  {https://doi.org/10.1016/j.progsolidstchem.2009.11.007} {\bibfield  {journal}
  {\bibinfo  {journal} {Prog. Solid State Chem.}\ }\textbf {\bibinfo {volume}
  {37}},\ \bibinfo {pages} {206} (\bibinfo {year} {2009})}\BibitemShut
  {NoStop}%
\bibitem [{\citenamefont {Ji}\ \emph {et~al.}(2021)\citenamefont {Ji},
  \citenamefont {Ding}, \citenamefont {Guan}, \citenamefont {Wu}, \citenamefont
  {Qian}, \citenamefont {Cao}, \citenamefont {Li},\ and\ \citenamefont
  {Jin}}]{ji2021}%
  \BibitemOpen
  \bibfield  {author} {\bibinfo {author} {\bibfnamefont {X.}~\bibnamefont
  {Ji}}, \bibinfo {author} {\bibfnamefont {D.}~\bibnamefont {Ding}}, \bibinfo
  {author} {\bibfnamefont {X.}~\bibnamefont {Guan}}, \bibinfo {author}
  {\bibfnamefont {C.}~\bibnamefont {Wu}}, \bibinfo {author} {\bibfnamefont
  {H.}~\bibnamefont {Qian}}, \bibinfo {author} {\bibfnamefont {J.}~\bibnamefont
  {Cao}}, \bibinfo {author} {\bibfnamefont {J.}~\bibnamefont {Li}},\ and\
  \bibinfo {author} {\bibfnamefont {C.}~\bibnamefont {Jin}},\ }\bibfield
  {title} {\bibinfo {title} {Interlayer {{Coupling Dependent Discrete H}}
  {$\rightarrow$} {{T'}} {{Phase Transition}} in {{Lithium Intercalated Bilayer
  Molybdenum Disulfide}}},\ }\href {https://doi.org/10.1021/acsnano.1c05332}
  {\bibfield  {journal} {\bibinfo  {journal} {ACS Nano}\ }\textbf {\bibinfo
  {volume} {15}},\ \bibinfo {pages} {15039} (\bibinfo {year}
  {2021})}\BibitemShut {NoStop}%
\bibitem [{\citenamefont {Xu}\ \emph {et~al.}(2024)\citenamefont {Xu},
  \citenamefont {Williams}, \citenamefont {Lee}, \citenamefont {Huang},
  \citenamefont {Siddique}, \citenamefont {Singer},\ and\ \citenamefont
  {Cha}}]{xuLi2024}%
  \BibitemOpen
  \bibfield  {author} {\bibinfo {author} {\bibfnamefont {S.}~\bibnamefont
  {Xu}}, \bibinfo {author} {\bibfnamefont {N.~L.}\ \bibnamefont {Williams}},
  \bibinfo {author} {\bibfnamefont {S.}~\bibnamefont {Lee}}, \bibinfo {author}
  {\bibfnamefont {J.~J.}\ \bibnamefont {Huang}}, \bibinfo {author}
  {\bibfnamefont {S.}~\bibnamefont {Siddique}}, \bibinfo {author}
  {\bibfnamefont {A.}~\bibnamefont {Singer}},\ and\ \bibinfo {author}
  {\bibfnamefont {J.~J.}\ \bibnamefont {Cha}},\ }\bibfield  {title} {\bibinfo
  {title} {Lithiation in {{2H-MoTe{\textsubscript{2}} Nanoflakes}}},\ }\href
  {https://doi.org/10.1021/acs.chemmater.4c01517} {\bibfield  {journal}
  {\bibinfo  {journal} {Chem. Mater.}\ }\textbf {\bibinfo {volume} {36}},\
  \bibinfo {pages} {10125} (\bibinfo {year} {2024})}\BibitemShut {NoStop}%
\bibitem [{\citenamefont {Huempfner}\ \emph {et~al.}(2022)\citenamefont
  {Huempfner}, \citenamefont {Otto}, \citenamefont {Fritz},\ and\ \citenamefont
  {Forker}}]{huempfner2022}%
  \BibitemOpen
  \bibfield  {author} {\bibinfo {author} {\bibfnamefont {T.}~\bibnamefont
  {Huempfner}}, \bibinfo {author} {\bibfnamefont {F.}~\bibnamefont {Otto}},
  \bibinfo {author} {\bibfnamefont {T.}~\bibnamefont {Fritz}},\ and\ \bibinfo
  {author} {\bibfnamefont {R.}~\bibnamefont {Forker}},\ }\bibfield  {title}
  {\bibinfo {title} {{$\pi$} {{Band Folding}} and {{Interlayer Band Filling}}
  of {{Graphene}} upon {{Interface Potassium Intercalation}}},\ }\href
  {https://doi.org/10.1002/admi.202200585} {\bibfield  {journal} {\bibinfo
  {journal} {Adv. Mater. Interfaces}\ }\textbf {\bibinfo {volume} {9}},\
  \bibinfo {pages} {2200585} (\bibinfo {year} {2022})}\BibitemShut {NoStop}%
\bibitem [{\citenamefont {Huempfner}\ \emph {et~al.}(2023)\citenamefont
  {Huempfner}, \citenamefont {Otto}, \citenamefont {Forker}, \citenamefont
  {M{\"u}ller},\ and\ \citenamefont {Fritz}}]{huempfner2023}%
  \BibitemOpen
  \bibfield  {author} {\bibinfo {author} {\bibfnamefont {T.}~\bibnamefont
  {Huempfner}}, \bibinfo {author} {\bibfnamefont {F.}~\bibnamefont {Otto}},
  \bibinfo {author} {\bibfnamefont {R.}~\bibnamefont {Forker}}, \bibinfo
  {author} {\bibfnamefont {P.}~\bibnamefont {M{\"u}ller}},\ and\ \bibinfo
  {author} {\bibfnamefont {T.}~\bibnamefont {Fritz}},\ }\bibfield  {title}
  {\bibinfo {title} {Superconductivity of {{K-Intercalated Epitaxial Bilayer
  Graphene}}},\ }\href {https://doi.org/10.1002/admi.202300014} {\bibfield
  {journal} {\bibinfo  {journal} {Adv. Mater. Interfaces}\ }\textbf {\bibinfo
  {volume} {10}},\ \bibinfo {pages} {2300014} (\bibinfo {year}
  {2023})}\BibitemShut {NoStop}%
\bibitem [{\citenamefont {Kim}\ \emph {et~al.}(2024)\citenamefont {Kim},
  \citenamefont {Sari}, \citenamefont {Chen}, \citenamefont {Rutt},
  \citenamefont {Ceder},\ and\ \citenamefont {Persson}}]{kim2024}%
  \BibitemOpen
  \bibfield  {author} {\bibinfo {author} {\bibfnamefont {J.}~\bibnamefont
  {Kim}}, \bibinfo {author} {\bibfnamefont {D.}~\bibnamefont {Sari}}, \bibinfo
  {author} {\bibfnamefont {Q.}~\bibnamefont {Chen}}, \bibinfo {author}
  {\bibfnamefont {A.}~\bibnamefont {Rutt}}, \bibinfo {author} {\bibfnamefont
  {G.}~\bibnamefont {Ceder}},\ and\ \bibinfo {author} {\bibfnamefont {K.~A.}\
  \bibnamefont {Persson}},\ }\bibfield  {title} {\bibinfo {title}
  {First-{{Principles}} and {{Experimental Investigation}} of
  {{ABO{\textsubscript{4}} Zircons}} as {{Calcium Intercalation Cathodes}}},\
  }\href {https://doi.org/10.1021/acs.chemmater.4c00062} {\bibfield  {journal}
  {\bibinfo  {journal} {Chem. Mater.}\ }\textbf {\bibinfo {volume} {36}},\
  \bibinfo {pages} {4444} (\bibinfo {year} {2024})}\BibitemShut {NoStop}%
\bibitem [{\citenamefont {Li}\ \emph {et~al.}(2024)\citenamefont {Li},
  \citenamefont {D{\"o}hn}, \citenamefont {Chen}, \citenamefont {Dillenz},
  \citenamefont {Sotoudeh}, \citenamefont {Pickup}, \citenamefont {Luo},
  \citenamefont {Parmenter}, \citenamefont {Arbiol}, \citenamefont
  {Alfredsson}, \citenamefont {Chadwick}, \citenamefont {Gro{\ss}},
  \citenamefont {Zarrabeitia},\ and\ \citenamefont {Ganin}}]{liK2024}%
  \BibitemOpen
  \bibfield  {author} {\bibinfo {author} {\bibfnamefont {W.}~\bibnamefont
  {Li}}, \bibinfo {author} {\bibfnamefont {J.}~\bibnamefont {D{\"o}hn}},
  \bibinfo {author} {\bibfnamefont {J.}~\bibnamefont {Chen}}, \bibinfo {author}
  {\bibfnamefont {M.}~\bibnamefont {Dillenz}}, \bibinfo {author} {\bibfnamefont
  {M.}~\bibnamefont {Sotoudeh}}, \bibinfo {author} {\bibfnamefont {D.~M.}\
  \bibnamefont {Pickup}}, \bibinfo {author} {\bibfnamefont {S.}~\bibnamefont
  {Luo}}, \bibinfo {author} {\bibfnamefont {R.}~\bibnamefont {Parmenter}},
  \bibinfo {author} {\bibfnamefont {J.}~\bibnamefont {Arbiol}}, \bibinfo
  {author} {\bibfnamefont {M.}~\bibnamefont {Alfredsson}}, \bibinfo {author}
  {\bibfnamefont {A.~V.}\ \bibnamefont {Chadwick}}, \bibinfo {author}
  {\bibfnamefont {A.}~\bibnamefont {Gro{\ss}}}, \bibinfo {author}
  {\bibfnamefont {M.}~\bibnamefont {Zarrabeitia}},\ and\ \bibinfo {author}
  {\bibfnamefont {A.~Y.}\ \bibnamefont {Ganin}},\ }\bibfield  {title} {\bibinfo
  {title} {Reversible {{K-ion}} intercalation in {{CrSe{\textsubscript{2}}}}
  cathodes for potassium-ion batteries: Combined operando {{PXRD}} and {{DFT}}
  studies},\ }\href {https://doi.org/10.1039/D4TA05114A} {\bibfield  {journal}
  {\bibinfo  {journal} {J. Mater. Chem. A}\ }\textbf {\bibinfo {volume} {12}},\
  \bibinfo {pages} {31276} (\bibinfo {year} {2024})}\BibitemShut {NoStop}%
\bibitem [{\citenamefont {Liu}\ \emph {et~al.}(2017)\citenamefont {Liu},
  \citenamefont {Wang}, \citenamefont {Li}, \citenamefont {Xie}, \citenamefont
  {Chen}, \citenamefont {Chen},\ and\ \citenamefont {Liu}}]{liuCu2017}%
  \BibitemOpen
  \bibfield  {author} {\bibinfo {author} {\bibfnamefont {R.}~\bibnamefont
  {Liu}}, \bibinfo {author} {\bibfnamefont {C.}~\bibnamefont {Wang}}, \bibinfo
  {author} {\bibfnamefont {Y.}~\bibnamefont {Li}}, \bibinfo {author}
  {\bibfnamefont {Y.}~\bibnamefont {Xie}}, \bibinfo {author} {\bibfnamefont
  {Q.}~\bibnamefont {Chen}}, \bibinfo {author} {\bibfnamefont {Z.}~\bibnamefont
  {Chen}},\ and\ \bibinfo {author} {\bibfnamefont {Q.}~\bibnamefont {Liu}},\
  }\bibfield  {title} {\bibinfo {title} {Intercalating copper into layered
  {{TaS{\textsubscript{2}}}} van der {{Waals}} gaps},\ }\href
  {https://doi.org/10.1039/C7RA08630J} {\bibfield  {journal} {\bibinfo
  {journal} {RSC Adv.}\ }\textbf {\bibinfo {volume} {7}},\ \bibinfo {pages}
  {46699} (\bibinfo {year} {2017})}\BibitemShut {NoStop}%
\bibitem [{\citenamefont {Nong}\ \emph {et~al.}(2025)\citenamefont {Nong},
  \citenamefont {Tan}, \citenamefont {Sun}, \citenamefont {Zhang},
  \citenamefont {Gu}, \citenamefont {Wei}, \citenamefont {Wang}, \citenamefont
  {Zhang}, \citenamefont {Wu}, \citenamefont {Zou},\ and\ \citenamefont
  {Liu}}]{nongCu2025}%
  \BibitemOpen
  \bibfield  {author} {\bibinfo {author} {\bibfnamefont {H.}~\bibnamefont
  {Nong}}, \bibinfo {author} {\bibfnamefont {J.}~\bibnamefont {Tan}}, \bibinfo
  {author} {\bibfnamefont {Y.}~\bibnamefont {Sun}}, \bibinfo {author}
  {\bibfnamefont {R.}~\bibnamefont {Zhang}}, \bibinfo {author} {\bibfnamefont
  {Y.}~\bibnamefont {Gu}}, \bibinfo {author} {\bibfnamefont {Q.}~\bibnamefont
  {Wei}}, \bibinfo {author} {\bibfnamefont {J.}~\bibnamefont {Wang}}, \bibinfo
  {author} {\bibfnamefont {Y.}~\bibnamefont {Zhang}}, \bibinfo {author}
  {\bibfnamefont {Q.}~\bibnamefont {Wu}}, \bibinfo {author} {\bibfnamefont
  {X.}~\bibnamefont {Zou}},\ and\ \bibinfo {author} {\bibfnamefont
  {B.}~\bibnamefont {Liu}},\ }\bibfield  {title} {\bibinfo {title} {Cu
  {{Intercalation-Stabilized 1T'}} {{MoS{\textsubscript{2}}}} with {{Electrical
  Insulating Behavior}}},\ }\href {https://doi.org/10.1021/jacs.4c14945}
  {\bibfield  {journal} {\bibinfo  {journal} {J. Am. Chem. Soc.}\ }\textbf
  {\bibinfo {volume} {147}},\ \bibinfo {pages} {9242} (\bibinfo {year}
  {2025})}\BibitemShut {NoStop}%
\bibitem [{\citenamefont {Sch{\"o}lzel}\ \emph {et~al.}(2024)\citenamefont
  {Sch{\"o}lzel}, \citenamefont {Richter}, \citenamefont {Unigarro},
  \citenamefont {Wolff}, \citenamefont {Schwarz}, \citenamefont {Sch{\"u}tze},
  \citenamefont {R{\"o}sch}, \citenamefont {Gemming}, \citenamefont {Seyller},\
  and\ \citenamefont {Sch{\"a}dlich}}]{scholzel2024}%
  \BibitemOpen
  \bibfield  {author} {\bibinfo {author} {\bibfnamefont {F.}~\bibnamefont
  {Sch{\"o}lzel}}, \bibinfo {author} {\bibfnamefont {P.}~\bibnamefont
  {Richter}}, \bibinfo {author} {\bibfnamefont {A.~D.~P.}\ \bibnamefont
  {Unigarro}}, \bibinfo {author} {\bibfnamefont {S.}~\bibnamefont {Wolff}},
  \bibinfo {author} {\bibfnamefont {H.}~\bibnamefont {Schwarz}}, \bibinfo
  {author} {\bibfnamefont {A.}~\bibnamefont {Sch{\"u}tze}}, \bibinfo {author}
  {\bibfnamefont {N.}~\bibnamefont {R{\"o}sch}}, \bibinfo {author}
  {\bibfnamefont {S.}~\bibnamefont {Gemming}}, \bibinfo {author} {\bibfnamefont
  {T.}~\bibnamefont {Seyller}},\ and\ \bibinfo {author} {\bibfnamefont
  {P.}~\bibnamefont {Sch{\"a}dlich}},\ }\bibfield  {title} {\bibinfo {title}
  {Large-{{Area Lead Monolayers}} under {{Cover}}: {{Intercalation}},
  {{Doping}}, and {{Phase Transformation}}},\ }\href
  {https://doi.org/10.1002/sstr.202400338} {\bibfield  {journal} {\bibinfo
  {journal} {Small Struct.}\ }\textbf {\bibinfo {volume} {6}},\ \bibinfo
  {pages} {2400338} (\bibinfo {year} {2024})}\BibitemShut {NoStop}%
\bibitem [{\citenamefont {Zhang}\ \emph {et~al.}(2020)\citenamefont {Zhang},
  \citenamefont {Niu}, \citenamefont {Tan}, \citenamefont {Deng}, \citenamefont
  {Jin}, \citenamefont {Zeng}, \citenamefont {Xu},\ and\ \citenamefont
  {Zhu}}]{zhangK2020}%
  \BibitemOpen
  \bibfield  {author} {\bibinfo {author} {\bibfnamefont {Y.}~\bibnamefont
  {Zhang}}, \bibinfo {author} {\bibfnamefont {X.}~\bibnamefont {Niu}}, \bibinfo
  {author} {\bibfnamefont {L.}~\bibnamefont {Tan}}, \bibinfo {author}
  {\bibfnamefont {L.}~\bibnamefont {Deng}}, \bibinfo {author} {\bibfnamefont
  {S.}~\bibnamefont {Jin}}, \bibinfo {author} {\bibfnamefont {L.}~\bibnamefont
  {Zeng}}, \bibinfo {author} {\bibfnamefont {H.}~\bibnamefont {Xu}},\ and\
  \bibinfo {author} {\bibfnamefont {Y.}~\bibnamefont {Zhu}},\ }\bibfield
  {title} {\bibinfo {title}
  {K{\textsubscript{0.83}}{{V{\textsubscript{2}}O{\textsubscript{5}}}}: {{A New
  Layered Compound}} as a {{Stable Cathode Material}} for {{Potassium-Ion
  Batteries}}},\ }\href {https://doi.org/10.1021/acsami.9b22087} {\bibfield
  {journal} {\bibinfo  {journal} {ACS Appl. Mater. Interfaces}\ }\textbf
  {\bibinfo {volume} {12}},\ \bibinfo {pages} {9332} (\bibinfo {year}
  {2020})}\BibitemShut {NoStop}%
\bibitem [{\citenamefont {Tonti}\ \emph {et~al.}(2000)\citenamefont {Tonti},
  \citenamefont {Pettenkofer},\ and\ \citenamefont {Jaegermann}}]{tonti2000}%
  \BibitemOpen
  \bibfield  {author} {\bibinfo {author} {\bibfnamefont {D.}~\bibnamefont
  {Tonti}}, \bibinfo {author} {\bibfnamefont {C.}~\bibnamefont {Pettenkofer}},\
  and\ \bibinfo {author} {\bibfnamefont {W.}~\bibnamefont {Jaegermann}},\
  }\bibfield  {title} {\bibinfo {title} {In-situ photoelectron spectroscopy
  study of a {{TiS{\textsubscript{2}}}} thin film cathode in an operating
  {{Na}} intercalation electrochemical cell},\ }\href
  {https://doi.org/10.1007/BF02374066} {\bibfield  {journal} {\bibinfo
  {journal} {Ionics}\ }\textbf {\bibinfo {volume} {6}},\ \bibinfo {pages} {196}
  (\bibinfo {year} {2000})}\BibitemShut {NoStop}%
\bibitem [{\citenamefont {Toyama}\ \emph {et~al.}(2022)\citenamefont {Toyama},
  \citenamefont {Akiyama}, \citenamefont {Ichinokura}, \citenamefont
  {Hashizume}, \citenamefont {Iimori}, \citenamefont {Endo}, \citenamefont
  {Hobara}, \citenamefont {Matsui}, \citenamefont {Horii}, \citenamefont
  {Sato}, \citenamefont {Hirahara}, \citenamefont {Komori},\ and\ \citenamefont
  {Hasegawa}}]{toyama2022}%
  \BibitemOpen
  \bibfield  {author} {\bibinfo {author} {\bibfnamefont {H.}~\bibnamefont
  {Toyama}}, \bibinfo {author} {\bibfnamefont {R.}~\bibnamefont {Akiyama}},
  \bibinfo {author} {\bibfnamefont {S.}~\bibnamefont {Ichinokura}}, \bibinfo
  {author} {\bibfnamefont {M.}~\bibnamefont {Hashizume}}, \bibinfo {author}
  {\bibfnamefont {T.}~\bibnamefont {Iimori}}, \bibinfo {author} {\bibfnamefont
  {Y.}~\bibnamefont {Endo}}, \bibinfo {author} {\bibfnamefont {R.}~\bibnamefont
  {Hobara}}, \bibinfo {author} {\bibfnamefont {T.}~\bibnamefont {Matsui}},
  \bibinfo {author} {\bibfnamefont {K.}~\bibnamefont {Horii}}, \bibinfo
  {author} {\bibfnamefont {S.}~\bibnamefont {Sato}}, \bibinfo {author}
  {\bibfnamefont {T.}~\bibnamefont {Hirahara}}, \bibinfo {author}
  {\bibfnamefont {F.}~\bibnamefont {Komori}},\ and\ \bibinfo {author}
  {\bibfnamefont {S.}~\bibnamefont {Hasegawa}},\ }\bibfield  {title} {\bibinfo
  {title} {Two-{{Dimensional Superconductivity}} of {{Ca-Intercalated
  Graphene}} on {{SiC}}: {{Vital Role}} of the {{Interface}} between
  {{Monolayer Graphene}} and the {{Substrate}}},\ }\href
  {https://doi.org/10.1021/acsnano.1c11161} {\bibfield  {journal} {\bibinfo
  {journal} {ACS Nano}\ }\textbf {\bibinfo {volume} {16}},\ \bibinfo {pages}
  {3582} (\bibinfo {year} {2022})}\BibitemShut {NoStop}%
\bibitem [{\citenamefont {Wang}\ \emph {et~al.}(2015)\citenamefont {Wang},
  \citenamefont {Shen}, \citenamefont {Gao}, \citenamefont {Wang},
  \citenamefont {Yu},\ and\ \citenamefont {Chen}}]{wangNaAl2015}%
  \BibitemOpen
  \bibfield  {author} {\bibinfo {author} {\bibfnamefont {X.}~\bibnamefont
  {Wang}}, \bibinfo {author} {\bibfnamefont {X.}~\bibnamefont {Shen}}, \bibinfo
  {author} {\bibfnamefont {Y.}~\bibnamefont {Gao}}, \bibinfo {author}
  {\bibfnamefont {Z.}~\bibnamefont {Wang}}, \bibinfo {author} {\bibfnamefont
  {R.}~\bibnamefont {Yu}},\ and\ \bibinfo {author} {\bibfnamefont
  {L.}~\bibnamefont {Chen}},\ }\bibfield  {title} {\bibinfo {title}
  {Atomic-{{Scale Recognition}} of {{Surface Structure}} and {{Intercalation
  Mechanism}} of {{Ti{\textsubscript{3}}C{\textsubscript{2}}X}}},\ }\href
  {https://doi.org/10.1021/ja512820k} {\bibfield  {journal} {\bibinfo
  {journal} {J. Am. Chem. Soc.}\ }\textbf {\bibinfo {volume} {137}},\ \bibinfo
  {pages} {2715} (\bibinfo {year} {2015})}\BibitemShut {NoStop}%
\bibitem [{\citenamefont {Naik}\ \emph {et~al.}(2021)\citenamefont {Naik},
  \citenamefont {Kalaiarasan}, \citenamefont {Nath}, \citenamefont {Sarangi},
  \citenamefont {Sahu}, \citenamefont {Samal}, \citenamefont {Biswal},\ and\
  \citenamefont {Samal}}]{naik2021}%
  \BibitemOpen
  \bibfield  {author} {\bibinfo {author} {\bibfnamefont {S.}~\bibnamefont
  {Naik}}, \bibinfo {author} {\bibfnamefont {S.}~\bibnamefont {Kalaiarasan}},
  \bibinfo {author} {\bibfnamefont {R.~C.}\ \bibnamefont {Nath}}, \bibinfo
  {author} {\bibfnamefont {S.~N.}\ \bibnamefont {Sarangi}}, \bibinfo {author}
  {\bibfnamefont {A.~K.}\ \bibnamefont {Sahu}}, \bibinfo {author}
  {\bibfnamefont {D.}~\bibnamefont {Samal}}, \bibinfo {author} {\bibfnamefont
  {H.~S.}\ \bibnamefont {Biswal}},\ and\ \bibinfo {author} {\bibfnamefont
  {S.~L.}\ \bibnamefont {Samal}},\ }\bibfield  {title} {\bibinfo {title}
  {Nominal {{Effect}} of {{Mg Intercalation}} on the {{Superconducting
  Properties}} of {{2H}}--{{NbSe{\textsubscript{2}}}}},\ }\href
  {https://doi.org/10.1021/acs.inorgchem.0c03545} {\bibfield  {journal}
  {\bibinfo  {journal} {Inorg. Chem.}\ }\textbf {\bibinfo {volume} {60}},\
  \bibinfo {pages} {4588} (\bibinfo {year} {2021})}\BibitemShut {NoStop}%
\bibitem [{\citenamefont {Meng}\ \emph {et~al.}(2020)\citenamefont {Meng},
  \citenamefont {Liu}, \citenamefont {Yang}, \citenamefont {Shi}, \citenamefont
  {Ge}, \citenamefont {Zhang}, \citenamefont {Ying}, \citenamefont {Wang},
  \citenamefont {Wang}, \citenamefont {Wu},\ and\ \citenamefont
  {Chen}}]{meng2020}%
  \BibitemOpen
  \bibfield  {author} {\bibinfo {author} {\bibfnamefont {F.~B.}\ \bibnamefont
  {Meng}}, \bibinfo {author} {\bibfnamefont {Z.}~\bibnamefont {Liu}}, \bibinfo
  {author} {\bibfnamefont {L.~X.}\ \bibnamefont {Yang}}, \bibinfo {author}
  {\bibfnamefont {M.~Z.}\ \bibnamefont {Shi}}, \bibinfo {author} {\bibfnamefont
  {B.~H.}\ \bibnamefont {Ge}}, \bibinfo {author} {\bibfnamefont
  {H.}~\bibnamefont {Zhang}}, \bibinfo {author} {\bibfnamefont {J.~J.}\
  \bibnamefont {Ying}}, \bibinfo {author} {\bibfnamefont {Z.~F.}\ \bibnamefont
  {Wang}}, \bibinfo {author} {\bibfnamefont {Z.~Y.}\ \bibnamefont {Wang}},
  \bibinfo {author} {\bibfnamefont {T.}~\bibnamefont {Wu}},\ and\ \bibinfo
  {author} {\bibfnamefont {X.~H.}\ \bibnamefont {Chen}},\ }\bibfield  {title}
  {\bibinfo {title} {Metal-insulator transition in organic ion intercalated
  {{VSe{\textsubscript{2}}}} induced by dimensional crossover},\ }\href
  {https://doi.org/10.1103/PhysRevB.102.165410} {\bibfield  {journal} {\bibinfo
   {journal} {Phys. Rev. B}\ }\textbf {\bibinfo {volume} {102}},\ \bibinfo
  {pages} {165410} (\bibinfo {year} {2020})}\BibitemShut {NoStop}%
\bibitem [{\citenamefont {Huang}\ \emph {et~al.}(2021)\citenamefont {Huang},
  \citenamefont {Xu}, \citenamefont {Zeng}, \citenamefont {Jiang},
  \citenamefont {Nie}, \citenamefont {Chen}, \citenamefont {Jiang},\ and\
  \citenamefont {Liu}}]{huangLi2021}%
  \BibitemOpen
  \bibfield  {author} {\bibinfo {author} {\bibfnamefont {X.}~\bibnamefont
  {Huang}}, \bibinfo {author} {\bibfnamefont {J.}~\bibnamefont {Xu}}, \bibinfo
  {author} {\bibfnamefont {R.}~\bibnamefont {Zeng}}, \bibinfo {author}
  {\bibfnamefont {Q.}~\bibnamefont {Jiang}}, \bibinfo {author} {\bibfnamefont
  {X.}~\bibnamefont {Nie}}, \bibinfo {author} {\bibfnamefont {C.}~\bibnamefont
  {Chen}}, \bibinfo {author} {\bibfnamefont {X.}~\bibnamefont {Jiang}},\ and\
  \bibinfo {author} {\bibfnamefont {J.-M.}\ \bibnamefont {Liu}},\ }\bibfield
  {title} {\bibinfo {title} {Li-ion intercalation enhanced ferromagnetism in
  van der {{Waals}} {{Fe{\textsubscript{3}}GeTe{\textsubscript{2}}}} bilayer},\
  }\href {https://doi.org/10.1063/5.0051882} {\bibfield  {journal} {\bibinfo
  {journal} {Appl. Phys. Lett.}\ }\textbf {\bibinfo {volume} {119}},\ \bibinfo
  {pages} {012405} (\bibinfo {year} {2021})}\BibitemShut {NoStop}%
\bibitem [{\citenamefont {Zhao}\ \emph {et~al.}(2020)\citenamefont {Zhao},
  \citenamefont {Song}, \citenamefont {Wang}, \citenamefont {{Riis-Jensen}},
  \citenamefont {Fu}, \citenamefont {Deng}, \citenamefont {Wan}, \citenamefont
  {Kang}, \citenamefont {Ning}, \citenamefont {Dan}, \citenamefont
  {Venkatesan}, \citenamefont {Liu}, \citenamefont {Zhou}, \citenamefont
  {Thygesen}, \citenamefont {Luo}, \citenamefont {Pennycook},\ and\
  \citenamefont {Loh}}]{zhaosoc2020}%
  \BibitemOpen
  \bibfield  {author} {\bibinfo {author} {\bibfnamefont {X.}~\bibnamefont
  {Zhao}}, \bibinfo {author} {\bibfnamefont {P.}~\bibnamefont {Song}}, \bibinfo
  {author} {\bibfnamefont {C.}~\bibnamefont {Wang}}, \bibinfo {author}
  {\bibfnamefont {A.~C.}\ \bibnamefont {{Riis-Jensen}}}, \bibinfo {author}
  {\bibfnamefont {W.}~\bibnamefont {Fu}}, \bibinfo {author} {\bibfnamefont
  {Y.}~\bibnamefont {Deng}}, \bibinfo {author} {\bibfnamefont {D.}~\bibnamefont
  {Wan}}, \bibinfo {author} {\bibfnamefont {L.}~\bibnamefont {Kang}}, \bibinfo
  {author} {\bibfnamefont {S.}~\bibnamefont {Ning}}, \bibinfo {author}
  {\bibfnamefont {J.}~\bibnamefont {Dan}}, \bibinfo {author} {\bibfnamefont
  {T.}~\bibnamefont {Venkatesan}}, \bibinfo {author} {\bibfnamefont
  {Z.}~\bibnamefont {Liu}}, \bibinfo {author} {\bibfnamefont {W.}~\bibnamefont
  {Zhou}}, \bibinfo {author} {\bibfnamefont {K.~S.}\ \bibnamefont {Thygesen}},
  \bibinfo {author} {\bibfnamefont {X.}~\bibnamefont {Luo}}, \bibinfo {author}
  {\bibfnamefont {S.~J.}\ \bibnamefont {Pennycook}},\ and\ \bibinfo {author}
  {\bibfnamefont {K.~P.}\ \bibnamefont {Loh}},\ }\bibfield  {title} {\bibinfo
  {title} {Engineering covalently bonded {{2D}} layered materials by
  self-intercalation},\ }\href {https://doi.org/10.1038/s41586-020-2241-9}
  {\bibfield  {journal} {\bibinfo  {journal} {Nature}\ }\textbf {\bibinfo
  {volume} {581}},\ \bibinfo {pages} {171} (\bibinfo {year}
  {2020})}\BibitemShut {NoStop}%
\bibitem [{\citenamefont {McGuire}\ \emph {et~al.}(2015)\citenamefont
  {McGuire}, \citenamefont {Dixit}, \citenamefont {Cooper},\ and\ \citenamefont
  {Sales}}]{mcguire2015}%
  \BibitemOpen
  \bibfield  {author} {\bibinfo {author} {\bibfnamefont {M.~A.}\ \bibnamefont
  {McGuire}}, \bibinfo {author} {\bibfnamefont {H.}~\bibnamefont {Dixit}},
  \bibinfo {author} {\bibfnamefont {V.~R.}\ \bibnamefont {Cooper}},\ and\
  \bibinfo {author} {\bibfnamefont {B.~C.}\ \bibnamefont {Sales}},\ }\bibfield
  {title} {\bibinfo {title} {Coupling of {{Crystal Structure}} and
  {{Magnetism}} in the {{Layered}}, {{Ferromagnetic Insulator
  CrI{\textsubscript{3}}}}},\ }\href {https://doi.org/10.1021/cm504242t}
  {\bibfield  {journal} {\bibinfo  {journal} {Chem. Mater.}\ }\textbf {\bibinfo
  {volume} {27}},\ \bibinfo {pages} {612} (\bibinfo {year} {2015})}\BibitemShut
  {NoStop}%
\bibitem [{\citenamefont {Ubrig}\ \emph {et~al.}(2019)\citenamefont {Ubrig},
  \citenamefont {Wang}, \citenamefont {Teyssier}, \citenamefont {Taniguchi},
  \citenamefont {Watanabe}, \citenamefont {Giannini}, \citenamefont
  {Morpurgo},\ and\ \citenamefont {Gibertini}}]{ubrig2019}%
  \BibitemOpen
  \bibfield  {author} {\bibinfo {author} {\bibfnamefont {N.}~\bibnamefont
  {Ubrig}}, \bibinfo {author} {\bibfnamefont {Z.}~\bibnamefont {Wang}},
  \bibinfo {author} {\bibfnamefont {J.}~\bibnamefont {Teyssier}}, \bibinfo
  {author} {\bibfnamefont {T.}~\bibnamefont {Taniguchi}}, \bibinfo {author}
  {\bibfnamefont {K.}~\bibnamefont {Watanabe}}, \bibinfo {author}
  {\bibfnamefont {E.}~\bibnamefont {Giannini}}, \bibinfo {author}
  {\bibfnamefont {A.~F.}\ \bibnamefont {Morpurgo}},\ and\ \bibinfo {author}
  {\bibfnamefont {M.}~\bibnamefont {Gibertini}},\ }\bibfield  {title} {\bibinfo
  {title} {Low-temperature monoclinic layer stacking in atomically thin
  {{CrI{\textsubscript{3}}}} crystals},\ }\href
  {https://doi.org/10.1088/2053-1583/ab4c64} {\bibfield  {journal} {\bibinfo
  {journal} {2D Mater.}\ }\textbf {\bibinfo {volume} {7}},\ \bibinfo {pages}
  {015007} (\bibinfo {year} {2019})}\BibitemShut {NoStop}%
\bibitem [{SMs()}]{SMs}%
  \BibitemOpen
  \bibfield  {title} {\bibinfo {title} {supplemental materials},\ }\href@noop
  {} {\ }\BibitemShut {NoStop}%
\bibitem [{\citenamefont {Wang}\ \emph {et~al.}(2018)\citenamefont {Wang},
  \citenamefont {{Guti{\'e}rrez-Lezama}}, \citenamefont {Ubrig}, \citenamefont
  {Kroner}, \citenamefont {Gibertini}, \citenamefont {Taniguchi}, \citenamefont
  {Watanabe}, \citenamefont {Imamo{\u g}lu}, \citenamefont {Giannini},\ and\
  \citenamefont {Morpurgo}}]{wangAFM2018}%
  \BibitemOpen
  \bibfield  {author} {\bibinfo {author} {\bibfnamefont {Z.}~\bibnamefont
  {Wang}}, \bibinfo {author} {\bibfnamefont {I.}~\bibnamefont
  {{Guti{\'e}rrez-Lezama}}}, \bibinfo {author} {\bibfnamefont {N.}~\bibnamefont
  {Ubrig}}, \bibinfo {author} {\bibfnamefont {M.}~\bibnamefont {Kroner}},
  \bibinfo {author} {\bibfnamefont {M.}~\bibnamefont {Gibertini}}, \bibinfo
  {author} {\bibfnamefont {T.}~\bibnamefont {Taniguchi}}, \bibinfo {author}
  {\bibfnamefont {K.}~\bibnamefont {Watanabe}}, \bibinfo {author}
  {\bibfnamefont {A.}~\bibnamefont {Imamo{\u g}lu}}, \bibinfo {author}
  {\bibfnamefont {E.}~\bibnamefont {Giannini}},\ and\ \bibinfo {author}
  {\bibfnamefont {A.~F.}\ \bibnamefont {Morpurgo}},\ }\bibfield  {title}
  {\bibinfo {title} {Very large tunneling magnetoresistance in layered magnetic
  semiconductor {{CrI{\textsubscript{3}}}}},\ }\href
  {https://doi.org/10.1038/s41467-018-04953-8} {\bibfield  {journal} {\bibinfo
  {journal} {Nat. Commun.}\ }\textbf {\bibinfo {volume} {9}},\ \bibinfo {pages}
  {2516} (\bibinfo {year} {2018})}\BibitemShut {NoStop}%
\bibitem [{\citenamefont {Anderson}(1959)}]{anderson1959}%
  \BibitemOpen
  \bibfield  {author} {\bibinfo {author} {\bibfnamefont {P.~W.}\ \bibnamefont
  {Anderson}},\ }\bibfield  {title} {\bibinfo {title} {New {{Approach}} to the
  {{Theory}} of {{Superexchange Interactions}}},\ }\href
  {https://doi.org/10.1103/PhysRev.115.2} {\bibfield  {journal} {\bibinfo
  {journal} {Phys. Rev.}\ }\textbf {\bibinfo {volume} {115}},\ \bibinfo {pages}
  {2} (\bibinfo {year} {1959})}\BibitemShut {NoStop}%
\bibitem [{\citenamefont {Goodenough}(1955)}]{goodenough1955}%
  \BibitemOpen
  \bibfield  {author} {\bibinfo {author} {\bibfnamefont {J.~B.}\ \bibnamefont
  {Goodenough}},\ }\bibfield  {title} {\bibinfo {title} {Theory of the {{Role}}
  of {{Covalence}} in the {{Perovskite-Type Manganites}} [{{La}},
  {{M}}({{II}})]{{MnO}}{\textsubscript{3}}},\ }\href
  {https://doi.org/10.1103/PhysRev.100.564} {\bibfield  {journal} {\bibinfo
  {journal} {Phys. Rev.}\ }\textbf {\bibinfo {volume} {100}},\ \bibinfo {pages}
  {564} (\bibinfo {year} {1955})}\BibitemShut {NoStop}%
\bibitem [{\citenamefont {Kanamori}(1960)}]{kanamori1960}%
  \BibitemOpen
  \bibfield  {author} {\bibinfo {author} {\bibfnamefont {J.}~\bibnamefont
  {Kanamori}},\ }\bibfield  {title} {\bibinfo {title} {Crystal {{Distortion}}
  in {{Magnetic Compounds}}},\ }\href {https://doi.org/10.1063/1.1984590}
  {\bibfield  {journal} {\bibinfo  {journal} {J. Appl. Phys.}\ }\textbf
  {\bibinfo {volume} {31}},\ \bibinfo {pages} {S14} (\bibinfo {year}
  {1960})}\BibitemShut {NoStop}%
\bibitem [{\citenamefont {Zener}(1951)}]{zener1951}%
  \BibitemOpen
  \bibfield  {author} {\bibinfo {author} {\bibfnamefont {C.}~\bibnamefont
  {Zener}},\ }\bibfield  {title} {\bibinfo {title} {Interaction between the
  d-{{Shells}} in the {{Transition Metals}}. {{II}}. {{Ferromagnetic
  Compounds}} of {{Manganese}} with {{Perovskite Structure}}},\ }\href
  {https://doi.org/10.1103/PhysRev.82.403} {\bibfield  {journal} {\bibinfo
  {journal} {Phys. Rev.}\ }\textbf {\bibinfo {volume} {82}},\ \bibinfo {pages}
  {403} (\bibinfo {year} {1951})}\BibitemShut {NoStop}%
\bibitem [{\citenamefont {Li}\ \emph {et~al.}(2023)\citenamefont {Li},
  \citenamefont {Zhang}, \citenamefont {You}, \citenamefont {Gu},\ and\
  \citenamefont {Su}}]{lisuperexchange2023}%
  \BibitemOpen
  \bibfield  {author} {\bibinfo {author} {\bibfnamefont {J.-W.}\ \bibnamefont
  {Li}}, \bibinfo {author} {\bibfnamefont {Z.}~\bibnamefont {Zhang}}, \bibinfo
  {author} {\bibfnamefont {J.-Y.}\ \bibnamefont {You}}, \bibinfo {author}
  {\bibfnamefont {B.}~\bibnamefont {Gu}},\ and\ \bibinfo {author}
  {\bibfnamefont {G.}~\bibnamefont {Su}},\ }\bibfield  {title} {\bibinfo
  {title} {Two-dimensional {{Heisenberg}} model with material-dependent
  superexchange interactions},\ }\href
  {https://doi.org/10.1103/PhysRevB.107.224411} {\bibfield  {journal} {\bibinfo
   {journal} {Phys. Rev. B}\ }\textbf {\bibinfo {volume} {107}},\ \bibinfo
  {pages} {224411} (\bibinfo {year} {2023})}\BibitemShut {NoStop}%
\bibitem [{\citenamefont {Yang}\ \emph {et~al.}(2025)\citenamefont {Yang},
  \citenamefont {Li}, \citenamefont {Yi}, \citenamefont {Li}, \citenamefont
  {You}, \citenamefont {Su},\ and\ \citenamefont {Gu}}]{yang2025}%
  \BibitemOpen
  \bibfield  {author} {\bibinfo {author} {\bibfnamefont {Q.-H.}\ \bibnamefont
  {Yang}}, \bibinfo {author} {\bibfnamefont {J.-W.}\ \bibnamefont {Li}},
  \bibinfo {author} {\bibfnamefont {X.-W.}\ \bibnamefont {Yi}}, \bibinfo
  {author} {\bibfnamefont {X.}~\bibnamefont {Li}}, \bibinfo {author}
  {\bibfnamefont {J.-Y.}\ \bibnamefont {You}}, \bibinfo {author} {\bibfnamefont
  {G.}~\bibnamefont {Su}},\ and\ \bibinfo {author} {\bibfnamefont
  {B.}~\bibnamefont {Gu}},\ }\bibfield  {title} {\bibinfo {title} {Enhancement
  of temperature of the quantum anomalous {{Hall}} effect in two-dimensional
  germanene/magnetic-semiconductor heterostructures},\ }\href
  {https://doi.org/10.1103/PhysRevB.111.184422} {\bibfield  {journal} {\bibinfo
   {journal} {Phys. Rev. B}\ }\textbf {\bibinfo {volume} {111}},\ \bibinfo
  {pages} {184422} (\bibinfo {year} {2025})}\BibitemShut {NoStop}%
\bibitem [{\citenamefont {Schrieffer}\ and\ \citenamefont
  {Wolff}(1966)}]{schrieffer1966}%
  \BibitemOpen
  \bibfield  {author} {\bibinfo {author} {\bibfnamefont {J.~R.}\ \bibnamefont
  {Schrieffer}}\ and\ \bibinfo {author} {\bibfnamefont {P.~A.}\ \bibnamefont
  {Wolff}},\ }\bibfield  {title} {\bibinfo {title} {Relation between the
  {{Anderson}} and {{Kondo Hamiltonians}}},\ }\href
  {https://doi.org/10.1103/PhysRev.149.491} {\bibfield  {journal} {\bibinfo
  {journal} {Phys. Rev.}\ }\textbf {\bibinfo {volume} {149}},\ \bibinfo {pages}
  {491} (\bibinfo {year} {1966})}\BibitemShut {NoStop}%
\bibitem [{\citenamefont {Mostofi}\ \emph {et~al.}(2008)\citenamefont
  {Mostofi}, \citenamefont {Yates}, \citenamefont {Lee}, \citenamefont {Souza},
  \citenamefont {Vanderbilt},\ and\ \citenamefont {Marzari}}]{Wannier902008}%
  \BibitemOpen
  \bibfield  {author} {\bibinfo {author} {\bibfnamefont {A.~A.}\ \bibnamefont
  {Mostofi}}, \bibinfo {author} {\bibfnamefont {J.~R.}\ \bibnamefont {Yates}},
  \bibinfo {author} {\bibfnamefont {Y.-S.}\ \bibnamefont {Lee}}, \bibinfo
  {author} {\bibfnamefont {I.}~\bibnamefont {Souza}}, \bibinfo {author}
  {\bibfnamefont {D.}~\bibnamefont {Vanderbilt}},\ and\ \bibinfo {author}
  {\bibfnamefont {N.}~\bibnamefont {Marzari}},\ }\bibfield  {title} {\bibinfo
  {title} {Wannier90: {{A}} tool for obtaining maximally-localised {{Wannier}}
  functions},\ }\href {https://doi.org/10.1016/j.cpc.2007.11.016} {\bibfield
  {journal} {\bibinfo  {journal} {Comput. Phys. Commun.}\ }\textbf {\bibinfo
  {volume} {178}},\ \bibinfo {pages} {685} (\bibinfo {year}
  {2008})}\BibitemShut {NoStop}%
\bibitem [{\citenamefont {Mostofi}\ \emph {et~al.}(2014)\citenamefont
  {Mostofi}, \citenamefont {Yates}, \citenamefont {Pizzi}, \citenamefont {Lee},
  \citenamefont {Souza}, \citenamefont {Vanderbilt},\ and\ \citenamefont
  {Marzari}}]{Wannier902014}%
  \BibitemOpen
  \bibfield  {author} {\bibinfo {author} {\bibfnamefont {A.~A.}\ \bibnamefont
  {Mostofi}}, \bibinfo {author} {\bibfnamefont {J.~R.}\ \bibnamefont {Yates}},
  \bibinfo {author} {\bibfnamefont {G.}~\bibnamefont {Pizzi}}, \bibinfo
  {author} {\bibfnamefont {Y.-S.}\ \bibnamefont {Lee}}, \bibinfo {author}
  {\bibfnamefont {I.}~\bibnamefont {Souza}}, \bibinfo {author} {\bibfnamefont
  {D.}~\bibnamefont {Vanderbilt}},\ and\ \bibinfo {author} {\bibfnamefont
  {N.}~\bibnamefont {Marzari}},\ }\bibfield  {title} {\bibinfo {title} {An
  updated version of wannier90: {{A}} tool for obtaining maximally-localised
  {{Wannier}} functions},\ }\href {https://doi.org/10.1016/j.cpc.2014.05.003}
  {\bibfield  {journal} {\bibinfo  {journal} {Comput. Phys. Commun.}\ }\textbf
  {\bibinfo {volume} {185}},\ \bibinfo {pages} {2309} (\bibinfo {year}
  {2014})}\BibitemShut {NoStop}%
\bibitem [{\citenamefont {Lyu}\ \emph {et~al.}(2022)\citenamefont {Lyu},
  \citenamefont {Zhang}, \citenamefont {You}, \citenamefont {Yan},\ and\
  \citenamefont {Su}}]{lyu2022}%
  \BibitemOpen
  \bibfield  {author} {\bibinfo {author} {\bibfnamefont {H.-Y.}\ \bibnamefont
  {Lyu}}, \bibinfo {author} {\bibfnamefont {Z.}~\bibnamefont {Zhang}}, \bibinfo
  {author} {\bibfnamefont {J.-Y.}\ \bibnamefont {You}}, \bibinfo {author}
  {\bibfnamefont {Q.-B.}\ \bibnamefont {Yan}},\ and\ \bibinfo {author}
  {\bibfnamefont {G.}~\bibnamefont {Su}},\ }\bibfield  {title} {\bibinfo
  {title} {Two-{{Dimensional Intercalating Multiferroics}} with {{Strong
  Magnetoelectric Coupling}}},\ }\href
  {https://doi.org/10.1021/acs.jpclett.2c03169} {\bibfield  {journal} {\bibinfo
   {journal} {J. Phys. Chem. Lett.}\ }\textbf {\bibinfo {volume} {13}},\
  \bibinfo {pages} {11405} (\bibinfo {year} {2022})}\BibitemShut {NoStop}%
\bibitem [{\citenamefont {Petrov}\ \emph {et~al.}(2021)\citenamefont {Petrov},
  \citenamefont {Ernst}, \citenamefont {Menshchikova},\ and\ \citenamefont
  {Chulkov}}]{petrov2021}%
  \BibitemOpen
  \bibfield  {author} {\bibinfo {author} {\bibfnamefont {E.~K.}\ \bibnamefont
  {Petrov}}, \bibinfo {author} {\bibfnamefont {A.}~\bibnamefont {Ernst}},
  \bibinfo {author} {\bibfnamefont {T.~V.}\ \bibnamefont {Menshchikova}},\ and\
  \bibinfo {author} {\bibfnamefont {E.~V.}\ \bibnamefont {Chulkov}},\
  }\bibfield  {title} {\bibinfo {title} {Intrinsic {{Magnetic Topological
  Insulator State Induced}} by the {{Jahn}}--{{Teller Effect}}},\ }\href
  {https://doi.org/10.1021/acs.jpclett.1c02396} {\bibfield  {journal} {\bibinfo
   {journal} {J. Phys. Chem. Lett.}\ }\textbf {\bibinfo {volume} {12}},\
  \bibinfo {pages} {9076} (\bibinfo {year} {2021})}\BibitemShut {NoStop}%
\bibitem [{\citenamefont {Maekawa}\ \emph {et~al.}(2004)\citenamefont
  {Maekawa}, \citenamefont {Tohyama}, \citenamefont {Barnes}, \citenamefont
  {Ishihara}, \citenamefont {Koshibae},\ and\ \citenamefont
  {Khaliullin}}]{maekawa2004}%
  \BibitemOpen
  \bibfield  {author} {\bibinfo {author} {\bibfnamefont {S.}~\bibnamefont
  {Maekawa}}, \bibinfo {author} {\bibfnamefont {T.}~\bibnamefont {Tohyama}},
  \bibinfo {author} {\bibfnamefont {S.~E.}\ \bibnamefont {Barnes}}, \bibinfo
  {author} {\bibfnamefont {S.}~\bibnamefont {Ishihara}}, \bibinfo {author}
  {\bibfnamefont {W.}~\bibnamefont {Koshibae}},\ and\ \bibinfo {author}
  {\bibfnamefont {G.}~\bibnamefont {Khaliullin}},\ }\href
  {https://doi.org/10.1007/978-3-662-09298-9} {\emph {\bibinfo {title} {Physics
  of {{Transition Metal Oxides}}}}},\ edited by\ \bibinfo {editor}
  {\bibfnamefont {M.}~\bibnamefont {Cardona}}, \bibinfo {editor} {\bibfnamefont
  {P.}~\bibnamefont {Fulde}}, \bibinfo {editor} {\bibfnamefont
  {K.}~\bibnamefont {Von~Klitzing}}, \bibinfo {editor} {\bibfnamefont
  {R.}~\bibnamefont {Merlin}}, \bibinfo {editor} {\bibfnamefont {H.-J.}\
  \bibnamefont {Queisser}},\ and\ \bibinfo {editor} {\bibfnamefont
  {H.}~\bibnamefont {St{\"o}rmer}},\ \bibinfo {series} {Springer {{Series}} in
  {{Solid-State Sciences}}}, Vol.\ \bibinfo {volume} {144}\ (\bibinfo
  {publisher} {Springer Berlin Heidelberg},\ \bibinfo {address} {Berlin,
  Heidelberg},\ \bibinfo {year} {2004})\BibitemShut {NoStop}%
\bibitem [{\citenamefont {Sivadas}\ \emph {et~al.}(2018)\citenamefont
  {Sivadas}, \citenamefont {Okamoto}, \citenamefont {Xu}, \citenamefont
  {Fennie},\ and\ \citenamefont {Xiao}}]{sivadas2018}%
  \BibitemOpen
  \bibfield  {author} {\bibinfo {author} {\bibfnamefont {N.}~\bibnamefont
  {Sivadas}}, \bibinfo {author} {\bibfnamefont {S.}~\bibnamefont {Okamoto}},
  \bibinfo {author} {\bibfnamefont {X.}~\bibnamefont {Xu}}, \bibinfo {author}
  {\bibfnamefont {{\relax Craig}.~J.}\ \bibnamefont {Fennie}},\ and\ \bibinfo
  {author} {\bibfnamefont {D.}~\bibnamefont {Xiao}},\ }\bibfield  {title}
  {\bibinfo {title} {Stacking-{{Dependent Magnetism}} in {{Bilayer
  CrI{\textsubscript{3}}}}},\ }\href
  {https://doi.org/10.1021/acs.nanolett.8b03321} {\bibfield  {journal}
  {\bibinfo  {journal} {Nano Lett.}\ }\textbf {\bibinfo {volume} {18}},\
  \bibinfo {pages} {7658} (\bibinfo {year} {2018})}\BibitemShut {NoStop}%
\end{thebibliography}


\end{document}